\documentstyle[aps,prb,multicol]{revtex}
\begin{document}
\title{Linear-response theory and lattice dynamics: a muffin-tin-orbital approach.}
\author{S. Y. Savrasov\cite{Lebedev}}
\address{Max-Planck-Institut f\"ur Festk\"orperforschung, Heisenbergstr.1, D--70569\\
Stuttgart, Germany.}
\date{\today }
\maketitle

\begin{abstract}
A detailed description of a method for calculating static linear-response
functions in the problem of lattice dynamics is presented. The method is
based on density functional theory and it uses linear muffin-tin orbitals as
a basis for representing first-order corrections to the one-electron wave
functions. This makes it possible to greatly facilitate the treatment of the
materials with localized orbitals. We derive variationally accurate
expressions for the dynamical matrix. We also show that large
incomplete-basis-set corrections to the first-order changes in the wave
functions exist and can be explicitly calculated. Some usefull hints on the $%
{\bf k}$-space integration for metals and the self-consistency problem at
long wavelengths are also given. As a test application we calculate bulk
phonon dispersions in $Si$ and find good agreement between our results and
experiments. As another application, we calculate lattice dynamics of the
transition-metal carbide $NbC$. The theory reproduces the major anomalies
found experimentally in its phonon dispersions. The theory also predicts an
anomalous behavior of the lowest transverse acoustic mode along the $(\xi
\xi 0)$ direction. Most of the calculated frequencies agree within a few
percent with those measured.
\end{abstract}


\begin{multicols}{2}

{\bf I. INTRODUCTION.}

\smallskip\ 

Response of electrons to a static external field is one of the important
characteristics of a solid which can be uniquely determined within density
functional theory (DFT) \cite{DFT}. The use of the local density
approximation (LDA) \cite{LDA,GAS} in the linear--response problem has by
now become a common and well established method for determining various
ground-state properties of real materials. These, first of all, include
static dielectric, structural, and vibrational properties of semiconductors
and insulators such as screening response to point charges and electric
fields, effective charges and dielectric constants, as well as whole phonon
spectra \cite{Louie,Baroni2,Blat,Gonze1,Krak}. {\it Ab initio} calculations
of the wave--vector dependent lattice dynamical properties for metallic
systems have also recently been performed \cite
{SSC,PHN,ZeinNb,EPI,UFN,Gironcoli}. Among them, transition metals, their
alloys and compound provide one of the fascinating areas in the study of
phonons in crystal lattice. This is because, in addition to the richness and
variety of structure of their phonon dispersion curves, these materials also
often exhibit lattice instabilities and relatively high--temperature
(8K--23K) superconductivity, in which phonons play a fundamental role. Here,
density functional based linear--response calculations provide an important
first step in studying such phenomena as electron--phonon interactions and
transport properties \cite{EPI} which describe phonon--limited electrical
and thermal resistivities, renormalization of specific heat (electron--mass
enhancement) as well as superconducting transition temperatures. These
properties are connected with the real low--energy excitation spectrum of a
metal.

Initially, two methods have been developed to deal with the perturbations
which break the periodicity of the original lattice. The first one, known as
direct or supercell approach \cite{Martin}, considers the perturbation with
wave vector ${\bf q}$ which is periodical in the new supercell structure.
This is possible if the wave vector is commensurate with the reciprocal
lattice of the supercell and only tractable computationally if the size of
the supercell is not large. This limits the applications to high-symmetry
wave vectors. The same technique can be applied to calculate the interplanar
force constants in direct space\cite{Chou}. The dynamical matrix is found
for any ${\bf q}$ using the Fourier transform provided the interatomic
interactions of a solid are of short range. Despite of the severe
computational--cost restrictions, the supercell approach has two important
advantages: (i) the electronic response and lattice dynamics can be studied
using programs for self--consistent band--structure calculations which are
standardly used in condense--matter physics, and, as a consequence, (ii),
all non--linear--response coefficients are easily obtained. Note, however,
that third--order non--linear coefficients can also be accessed within the
linear--response approach \cite{Gonze2} just like forces are found within
the density--functional total--energy method. We shall discuss this
statement in more detail later in this paper.

The second method to deal with the perturbations is known as the
perturbative approach. If the external perturbation is weak one can use
standard perturbation theory and expand the first--order corrections to the
one--electron wave functions in the unperturbed Bloch states of the original
crystal. Usually it is done by introducing the so--called
independent--particle polarizability function in terms of which the screened
perturbation is found by inverting the static dielectric matrix of a crystal 
\cite{LRT} . Previously, due to a rapid progress made in the microscopic
theory of the phonon spectra in free--electron--like metals through the
development and application of the pseudopotential technique, plane--wave
representation was used for all the relevant quantities in the calculation.
However, already in the case of covalent semiconductors with sufficiently
weak pseudopotential, the convergency of the polarizability with respect to
a number of plane waves becomes slow and there are only a few attempts to
compute phonon spectra within this framework \cite{Devreese}. The situation
becomes worse for materials with localized orbitals. The most
time--consuming step in this approach is connected with the problem of
summation over high--energy states which at least requires their
calculation. Another problem is connected with the inversion of a large
dielectric matrix.

The above mentioned drawbacks of the perturbative approach have been
circumvented using the solid--state generalization \cite{Baroni,Zein} of the
Sternheimer method \cite{Stern}. In this reformulated linear--response
method, the first--order corrections to the unperturbed wave functions are
found by solving the Sternheimer equation (which is the Schr\"odinger
equation to linear order) directly without using the expansions over
unperturbed states. This avoids the summation problem of the perturbation
theory. The screening of the external perturbation is calculated
self--consistently within DFT in close analogy with what is done in standard
band--structure calculations. This avoids the inversion problem. The present
formulation is thus very efficient computationally which is demonstrated by
an increasing number of applications to the problem of lattice dynamics in
recent years. \cite
{Baroni2,Blat,Gonze1,Krak,SSC,PHN,ZeinNb,EPI,UFN,Gironcoli}.

In order to solve the Sternheimer equation, one has to construct a
rapidly--convergent basis set for representing the first--order
perturbations. This is important because these corrections as well as the
unperturbed wave functions oscillate in the core region. In the
free--electron--like metals, broad--band semiconductors and insulators, this
problem can be eliminated by the pseudopotential approximation and the
majority of the applications performed so far use plane--wave basis sets 
\cite{Baroni2,Gonze1,Gironcoli}. Unfortunately, with decreasing bandwidth,
the plane--wave expansion of the pseudo wave functions converges more slowly
and it becomes less advantageous to employ pseudopotentials. Indeed, until
most recently \cite{SSC,PHN,ZeinNb,EPI,UFN,Gironcoli}, the literature
contains no {\it ab initio} calculations of phonon dispersions for
transition--metal systems.

In the present paper we describe an efficient all--electron generalization
of the linear--response approach introduced in Refs. \onlinecite{Baroni,Zein}
(A brief report of this work has been published already \cite{PHN}). The
first--order corrections are represented in terms of the muffin--tin (MT)
basis sets which greatly facilitate the treatment of localized valence wave
functions. While the approach developed in the paper is general and can be
applied to any kind of localized--orbital representation, we use the
linear--muffin--tin--orbital (LMTO) method \cite{OKA} as a framework of such
all--electron formulation.

There are two problems addressed in this paper which are connected with the
use of MT basis functions in a linear--response method. The first problem
concerns the construction of a variational solution of the Sternheimer
equation. This is necessary because the unperturbed energy bands and wave
functions are obtained within the LMTO\ method by applying the
Rayleigh--Ritz variational principle. They are, therefore, not exact
solutions of the one--electron Schr\"odinger equation. As first shown by
Pulay \cite{Pulay}, the use of variational solutions gives rise to the
incomplete--basis--set (IBS) corrections in force calculations. The IBS
corrections must be carefully accounted for to get accurate forces in the
LMTO\ method \cite{FORCE}. These corrections also exist and must be taken
into account when calculating the first--order changes in the wave functions
and the dynamical matrix within the linear--response approach. The other
problem is connected with the change in the basis functions due to the
perturbation. Since the one--electron wave function in the LMTO\ method is
represented by the expansion coefficients in the MT basis set, under static
external perturbation, such as the displacement of a nucleus from its
equilibrium position, the change in the wave function will be described by
both the change in the expansion coefficients and the change in the basis
set. The contribution from the change in the basis set is important because
the original basis set is tailored to the one--electron potential and must
therefore be reconstructed to account for the specifics of the perturbation.
It should be noted that this contribution is not taken into account in the
standard perturbation theory, where only the change in the expansion
coefficients is taken into account.

The rest of the paper is organized as follows. The variational formulation
of the linear--response approach is described in Section II. Implementation
of the theory in the framework of the LMTO method is described in Section
III. Application of the method to phonon spectra in $Si$ and $NbC$ is given
in Section IV. Section V concludes the paper.

\bigskip\ \ 

{\bf II. THEORY.}

\smallskip\ 

{\bf a. Density-functional linear response.}

\smallskip\ 

Within density functional theory, the problem of calculating the lattice
dynamics essentially amounts to finding the change in the electronic charge
density induced by the presence of a phonon with wave vector ${\bf q}$.
Consider a lattice with a few atoms in the unit cell given by the positions $%
{\bf R}+{\bf t}$, where ${\bf R}$ are the basis vectors and ${\bf t}$ are
the translation vectors. Suppose that the atoms are displaced from their
equilibrium positions by a small amount:

\begin{equation}
\delta {\bf t}_R=\delta {\bf A}_Rexp(+i{\bf qt})+\delta {\bf A}_R^{*}exp(-i%
{\bf qt}),  \label{2.0}
\end{equation}
where $\delta {\bf A}_R$ is a complex polarization vector and ${\bf q}$ is
the phonon wave vector. The presence of such displacement field changes the
bare Coulomb potential as follows

\begin{equation}
\tilde V_{ext}({\bf r})=\sum_{R,t}\frac{-Z_Re^2}{|{\bf r}-{\bf R}-{\bf t}%
-\delta {\bf t}_R|},  \label{2.1}
\end{equation}
where $Z_R$ are the nuclei charges. By expanding this expression to first
order in the displacements we obtain that the crystal is perturbed by the
static external field:

\begin{eqnarray}
\delta V_{ext}({\bf r}) &=&\sum_R\delta {\bf A}_R\sum_te^{+i{\bf qt}}{\bf %
\nabla }\frac{Z_Re^2}{|{\bf r}-{\bf R}-{\bf t}|}+  \nonumber \\
&&\sum_R\delta {\bf A}_R^{*}\sum_te^{-i{\bf qt}}{\bf \nabla }\frac{Z_Re^2}{|%
{\bf r}-{\bf R}-{\bf t}|},  \label{2.2}
\end{eqnarray}
which is represented as a superposition of two travelling waves with wave
vectors $+{\bf q}$ and $-{\bf q}$, i.e.

\begin{equation}
\delta V_{ext}({\bf r})=\sum_R\delta {\bf A}_R\frac{\delta ^{+}V_{ext}({\bf r%
})}{\delta {\bf R}}+\sum_R\delta {\bf A}_R^{*}\frac{\delta ^{-}V_{ext}({\bf r%
})}{\delta {\bf R}}.  \label{2.3}
\end{equation}
To shorten the notations, we will omit $\delta {\bf R}$ from this definition
and, therefore, $\delta V_{ext}=\sum \delta {\bf A}\delta ^{+}V_{ext}+\sum
\delta {\bf A}^{*}\delta ^{-}V_{ext}.$ Both components $\delta ^{+}V_{ext}$
and $\delta ^{-}V_{ext}$ are hermitian: $[\delta ^{\pm }V_{ext}]^{*}=\delta
^{\mp }V_{ext\text{ }}$and translate like Bloch waves in the original
crystal: $\delta ^{\pm }V_{ext}({\bf r}+{\bf t})=exp(\pm i{\bf qt})\delta
^{\pm }V_{ext}({\bf r})$ .

The first--order change in the charge density, $\delta \rho ,$ induced by
the perturbation (\ref{2.2}) is represented in the same form as (\ref{2.3}),
i.e. $\delta \rho =\sum \delta {\bf A}\delta ^{+}\rho +\sum \delta {\bf A}%
^{*}\delta ^{-}\rho \,$ and it is expressed in terms of the one-electron
wave functions $\psi _{{\bf k}j}$ and their first-order corrections $\delta
^{+}\psi _{{\bf k}j}$ and $\delta ^{-}\psi _{{\bf k}j}$ as follows

\begin{equation}
\delta ^{\pm }\rho =\sum_{{\bf k}j}f_{{\bf k}j}(\delta ^{\pm }\psi _{{\bf k}%
j}^{*}\psi _{{\bf k}j}+\psi _{{\bf k}j}^{*}\delta ^{\pm }\psi _{{\bf k}j}),
\label{2.4}
\end{equation}
where $f_{{\bf k}j}$ are the occupation numbers, ${\bf k}$ lies in the first
Brillouin zone, and $j$ is the band index. The first--order correction $%
\delta ^{\pm }\psi _{{\bf k}j}\equiv |\delta ^{\pm }{\bf k}j\rangle =(\delta
^{\mp }\psi _{{\bf k}j})^{*}$ is a Bloch wave with vector ${\bf k}\pm {\bf q}
$ and it is the solution of the Sternheimer equation, which is the
Schr\"odinger equation to linear order:

\begin{equation}
(-\nabla ^2+V_{eff}-\epsilon _{{\bf k}j})|\delta ^{\pm }{\bf k}j\rangle
+\delta ^{\pm }V_{eff}|{\bf k}j\rangle =0.  \label{2.5}
\end{equation}
Here, $V_{eff}$ is the effective DFT potential and $\delta ^{\pm }V_{eff}$
is the change in the potential which is the external perturbation screened
by the induced charge density:

\begin{equation}
\delta ^{\pm }V_{eff}=\delta ^{\pm }V_{ext}+e^2\int \frac{\delta ^{\pm }\rho 
}{|{\bf r}-{\bf r}^{\prime }|}+\frac{dV_{xc}}{d\rho }\delta ^{\pm }\rho ,
\label{2.6}
\end{equation}
where the exchange--correlation effects are taken into account in the local
density approximation. In Eq. (\ref{2.5}) we have dropped the term with the
first--order corrections to the one--electron energies: $\delta ^{\pm
}\epsilon _{{\bf k}j}=\langle {\bf k}j|\delta ^{\pm }V_{eff}|{\bf k}j\rangle
,$ which are equal to zero if ${\bf q}\neq 0.\,$The Eqs.(\ref{2.4})--(\ref
{2.6}) must be solved self-consistently, i.e. (i) one has to solve (\ref{2.5}%
) with the external perturbation $\delta ^{\pm }V_{ext}$, or the one
screened by some guessed $\delta ^{\pm }\rho $, (ii) find the induced charge
density according to (\ref{2.4}) and, (iii), find new $\delta ^{\pm }V_{eff}$
after (\ref{2.6}). Steps (i)--(iii) are repeated until input and output $%
\delta ^{\pm }\rho $ coincide within a required accuracy. This is much
analogously to finding the unperturbed quantities $\psi _{{\bf k}j}$, $\rho $
and $V_{eff}$ in standard band--structure calculations.

\smallskip\ 

{\bf b. Dynamical matrix.}

\smallskip\ 

We must now solve two problems in order to calculate the lattice dynamics:
to develop a method for solving the Sternheimer equation (\ref{2.5}), and to
find an expression for the dynamical matrix. The general strategy employed
in the following is to consider the dynamical matrix as a functional of the
first--order perturbations. Expanding $\delta ^{\pm }\psi _{{\bf k}j}$ in
terms of the MT-basis functions will lead, under the stationarity condition,
to a set of matrix equations which represent a variational solution to Eq. (%
\ref{2.5}).

The variational formulation of the linear--response problem is required
because the original states $\psi _{{\bf k}j}$ are normally found not as
exact but variational solutions of the one--electron Schr\"odinger equation.
In an all--electron method such as the LMTO method, the wave function is
expanded in terms of the basis set $|\chi _\alpha ^{{\bf k}}\rangle $:

\begin{equation}
\psi _{{\bf k}j}=\sum_\alpha |\chi _\alpha ^{{\bf k}}\rangle A_\alpha ^{{\bf %
k}j}.  \label{2.7}
\end{equation}
The total energy is then considered as a functional of only the expansion
coefficients $A_\alpha ^{{\bf k}j},$ which are found by applying the
Rayleigh--Ritz variational principle. This leads to the following matrix
eigenvalue problem:

\begin{equation}
\sum_\alpha \langle \chi _\beta ^{{\bf k}}|-\nabla ^2+V_{eff}-\epsilon _{%
{\bf k}j}|\chi _\alpha ^{{\bf k}}\rangle A_\alpha ^{{\bf k}j}=0,  \label{2.8}
\end{equation}
which, in particular, allows all non--spherical terms in the potential to be
taken explicitly into account .

In the problem of lattice dynamics the second--order change in the total
energy must be found. It is obtained by expanding the total energy with
respect to the change in the external potential (nuclei displacements) up to
second order, i.e. $E=E_0+\delta ^{(1)}E+\delta ^{(2)}E.$ The first-order
change $\delta ^{(1)}E$ vanishes if the lattice is in the equilibrium and
the second--order change is expressed via the dynamical matrix of a solid:

\begin{equation}
\delta ^{(2)}E=\frac 12\sum_{R^{\prime }\mu ^{\prime }R\mu }\Lambda
_{R^{\prime }R}^{\mu ^{\prime }\mu }({\bf q})\delta A_{R^{\prime }\mu
^{\prime }}\delta A_{R\mu }^{*}+c.c.,  \label{I1}
\end{equation}
where we assumed that the nuclear displacements have the form (\ref{2.0})
and where $\{\mu \}$ denote directions of the polarization. A formula for
dynamical matrix $\Lambda $ is obtained by varying a density-functional
expression for the total energy. It is given by

\begin{eqnarray}
\ &&\Lambda _{R^{\prime }R}^{\mu ^{\prime }\mu }({\bf q})\equiv \Lambda = 
\nonumber \\
&&\ \ \sum_{{\bf k}j}f_{{\bf k}j}\langle \delta ^{+}\delta ^{-}{\bf k}%
j+\delta ^{-}\delta ^{+}{\bf k}j|-\nabla ^2+V_{eff}-\epsilon _{{\bf k}j}|%
{\bf k}j\rangle +  \nonumber \\
&&\ \ \sum_{{\bf k}j}f_{{\bf k}j}2\langle \delta ^{+}{\bf k}j|-\nabla
^2+V_{eff}-\epsilon _{{\bf k}j}|\delta ^{+}{\bf k}j\rangle +  \nonumber \\
&&\ \ \int \delta ^{+}\rho \delta ^{-}V_{eff}+\int \delta ^{+}\rho \delta
^{-}V_{ext}+\int \rho \delta ^{+}\delta ^{-}V_{ext}.  \nonumber \\
&&  \label{2.9}
\end{eqnarray}
(We omit indexes $R^{\prime }\mu ^{\prime }R\mu $ for simplicity.) Here $%
|\delta ^{\pm }\delta ^{\mp }{\bf k}j\rangle \,$ denote second--order
changes of the wave functions, and $\delta ^{+}\delta ^{-}V_{ext}$ is the
second--order change in the bare Coulomb potential due to the displacements
of nuclei. The latter is given by

\begin{equation}
\delta ^{+}\delta ^{-}V_{ext}\equiv \frac{\delta ^{+}\delta ^{-}V_{ext}}{%
\delta R_{\mu ^{\prime }}^{^{\prime }}\delta R_\mu }=\delta _{R^{\prime
}R}\sum_t\nabla _{\mu ^{\prime }}\nabla _\mu \frac{Z_Re^2}{|{\bf r}-{\bf R}-%
{\bf t}|}.  \label{2.10}
\end{equation}
Note that, when expanding Eq. (\ref{2.1}), it is only necessary to keep the
terms periodical in the original lattice, because only these give rise to a
non--zero contribution to the integral with $\rho $. Consequently, all the
contributions with the phase factor $exp(\pm 2i{\bf qt})$ can be neglected.
The same is true for the second--order terms: $|\delta ^{\pm }\delta ^{\mp }%
{\bf k}j\rangle $ are the functions with wave vector ${\bf k}$ since they
contribute to the matrix elements with $|{\bf k}j\rangle $ only. (Therefore,
the operator $\delta ^{\pm }$ can be considered as a variation of a Bloch
wave, whereby $\pm {\bf q}$ gets added to its wave vector.)

The first term in expression (\ref{2.9}) is not zero if the unperturbed
states are approximate solutions found from the eigenvalue problem (\ref{2.8}%
). If, on the other hand, one neglects it, and performs a variation of $%
\Lambda $ with respect to the first--order corrections, the self--consistent
linear--response equations (\ref{2.5}), (\ref{2.6}) will be recovered. The
expression (\ref{2.9}) is variational with respect to the first--order
changes in the wave functions just like the unperturbed total energy is
variational with respect to the unperturbed states $|{\bf k}j\rangle .$ This
property of the density functional follows from the Hohenberg--Kohn--Sham
variational principle. Eq. (\ref{2.9}) is directly interpreted as the
electronic contribution to the dynamical matrix. (The part connected with
change in the bare Coulomb energy of the nuclei is evaluated trivially by
the Ewald technique.) Due to stationarity of this functional the calculation
of the dynamical matrix is very accurate: while the first--order changes in
the wave functions and the charge densities are only variationally accurate,
the error will be of second order with respect to the error in $|\delta
^{\pm }{\bf k}j\rangle .$ In particular, the convergence of the dynamical
matrix during the iterations towards self--consistency is much faster than
the convergence of the induced charge density. At its minimum, expression (%
\ref{2.9}) contains no second and third terms and it may therefore be
interpreted as the Hellmann--Feyman result (last two terms) plus
incomplete--basis--set correction [the first contribution]. The latter is
the analog of the ''Pulay force'' known from force calculations.

The discussed variational properties of the dynamical matrix are indeed not
unique and represent a particular case of the powerful ''$2n+1$'' theorem of
perturbation theory: the knowledge of the perturbations in the wave
functions up to $(n)$th order allows one to find the $(2n+1)$th correction
to the eigenenergy. A generalization to arbitrary order of perturbation
within the density functional theory as well as variational properties of
even derivatives of the total energy were discussed in Ref. %
\onlinecite{Gonze2}. A direct minimization of the dynamical matrix by the
conjugate-gradient method has also been demonstrated recently \cite{Gonze1}.
It worth to point out that the knowledge of the first--order corrections to
the wave functions allows us to consider changes in the total--energy up to 
{\em third}{\it \ }order, in the same way as the zeroth order unperturbed
states allow calculating such first--order derivatives as, for instance,
forces. Consequently, third--order unharmonicity constants, Gruneizen
parameters and other non--linear coefficients are, in principle, easily
accessed within the linear--response formalism.

\smallskip\ 

{\bf c. First-order corrections.}

\smallskip\ 

We now turn out to the construction of the basis functions which represent
the first--order perturbations. Since the unperturbed state is given by
expansion (\ref{2.7}), the first--order change $|\delta ^{\pm }{\bf k}%
j\rangle $ must include both the change $\delta ^{\pm }A_\alpha ^{{\bf k}j}$
in the expansion coefficients as well as the change $|\delta ^{\pm }\chi
_\alpha ^{{\bf k}}\rangle $ in the MT basis set, i.e.

\begin{equation}
|\delta ^{\pm }{\bf k}j\rangle =\sum_\alpha \{|\chi _\alpha ^{{\bf k}+{\bf q}%
}\rangle \delta ^{\pm }A_\alpha ^{{\bf k}j}+|\delta ^{\pm }\chi _\alpha ^{%
{\bf k}}\rangle A_\alpha ^{{\bf k}j}\}.  \label{2.11}
\end{equation}
Since $|\delta ^{\pm }{\bf k}j\rangle $ is a Bloch state with wave vector $%
{\bf k}\pm {\bf q}$ so are $|\chi _\alpha ^{{\bf k+q}}\rangle $ and $|\delta
^{\pm }\chi _\alpha ^{{\bf k}}\rangle $ . The first function is the original
linear MT orbital of wave vector ${\bf k}\pm {\bf q}$ and the second one is
the change in the MT orbital due to the movements of atoms. In Section III
we will give detailed formulae for the change in the basis functions. Here
we note that since the original basis $|\chi _\alpha ^{{\bf k}}\rangle $ is
a Bloch sum of atom--centered localized orbitals, the important contribution
to the change in the Bloch sum is connected with the {\em rigid} movement of
these orbitals due to the rigid movement of the potentials for displaced
atoms. The expansion (\ref{2.11}) is rapidly convergent because the basis $%
|\delta ^{\pm }\chi _\alpha ^{{\bf k}}\rangle $ can be tailored to the
perturbation just like the basis $|\chi _\alpha ^{{\bf k}}\rangle $ is
tailored to the unperturbed potential. Eq. (\ref{2.11}) can be interpreted
as an expansion of $|\delta ^{\pm }{\bf k}j\rangle $ in terms of $|\chi
_\alpha ^{{\bf k+q}}\rangle $ in the local coordinate system displaced with
the atom; the convergence with respect to the number of orbitals per atom
must be about the same as for the unperturbed state. This is in contrast
with the expression of the standard perturbation theory where, for the
expansion of $|\delta ^{\pm }{\bf k}j\rangle ,$ only the change in the
coefficients $A_\alpha ^{{\bf k}j}$ is taken into account [first
contribution to (\ref{2.11})].

We shall now consider the second--order changes $|\delta ^{\pm }\delta ^{\mp
}{\bf k}j\rangle $ which appear in expression (\ref{2.9}) for the
incomplete--basis--set corrections to the dynamical matrix. By performing
the variation of the expansion (\ref{2.7}) to second order we obtain:

\begin{eqnarray}
|\delta ^{\pm }\delta ^{\mp }{\bf k}j\rangle  &=&\sum_\alpha \{|\chi _\alpha
^{{\bf k}}\rangle \delta ^{\pm }\delta ^{\mp }A_\alpha ^{{\bf k}j}+|\delta
^{\mp }\chi _\alpha ^{{\bf k}\pm {\bf q}}\rangle \delta ^{\pm }A_\alpha ^{%
{\bf k}j}+  \nonumber \\
&&+|\delta ^{\pm }\chi _\alpha ^{{\bf k}\mp {\bf q}}\rangle \delta ^{\mp
}A_\alpha ^{{\bf k}j}+|\delta ^{\pm }\delta ^{\mp }\chi _\alpha ^{{\bf k}%
}\rangle A_\alpha ^{{\bf k}j}\},  \label{2.13}
\end{eqnarray}
where $\delta ^{\pm }\delta ^{\mp }A_\alpha ^{{\bf k}j}$ and $|\delta ^{\pm
}\delta ^{\mp }\chi _\alpha ^{{\bf k}}\rangle $ are the second--order
changes in the expansion coefficients and the basis functions respectively.
By inserting (\ref{2.13}) in the first term of (\ref{2.9}) one sees that the
second--order changes $\delta ^{\pm }\delta ^{\mp }A_\alpha ^{{\bf k}j}$ do
not contribute because they enter as coefficients to the unperturbed basis
functions and

\begin{equation}
\sum_\alpha \delta ^{\pm }\delta ^{\mp }A_\alpha ^{{\bf k}j}\langle \chi
_\alpha ^{{\bf k}}|H-\epsilon _{{\bf k}j}|{\bf k}j\rangle \equiv 0.
\label{2.14}
\end{equation}
(Here, $H=-\nabla ^2+V_{eff\text{.}})$

The absence of the coefficients $\delta ^{\pm }\delta ^{\mp }A_\alpha ^{{\bf %
k}j}$ in our formulation of the problem has an important consequence: since $%
|\delta ^{\pm }\delta ^{\mp }{\bf k}j\rangle $ has only the unknown
contribution from the first--order changes in $A_\alpha ^{{\bf k}j}$ and
since the Hilbert space $\{|\chi \rangle ,|\delta \chi \rangle \}$ of the
basis functions is fixed, we see that the variational freedom of the
functional (\ref{2.9}) is provided only by the coefficients $\delta ^{\pm
}A_\alpha ^{{\bf k}j}.$ This is again in close analogy to that in
band--structure calculations the variational freedom of the total energy is
provided only by the unperturbed coefficients $A_\alpha ^{{\bf k}j}$. In the
total--energy calculations this has the consequence when calculating the
forces: due to the stationarity condition the force formula does not
contains any first--order derivatives in $A_\alpha ^{{\bf k}j}$. In the
dynamical--matrix calculation this will have the same consequence when
calculating third--order non--linear coefficients: the corresponding
formulae will not contain any second-- and third--order derivatives of $%
A_\alpha ^{{\bf k}j}$ and, thus, can be explicitly evaluated from only the
knowledge of $\delta ^{\pm }A_\alpha ^{{\bf k}j}$. Note however, that
together with the matrix elements containing $|\chi \rangle ,|\delta \chi
\rangle ,|\delta ^{(2)}\chi \rangle $, a contribution from third--order
changes in the basis sets must be taken into account.

We shall now derive the equations for the first--order changes in the
expansion coefficients. This is done by minimization of $\Lambda $ with
respect to $\delta ^{\pm }A_\alpha ^{{\bf k}j}$. We obtain:

\begin{eqnarray}
&&\sum_\alpha \langle \chi _\beta ^{{\bf k}\pm {\bf q}}|H-\epsilon _{{\bf k}%
j}|\chi _\alpha ^{{\bf k}\pm {\bf q}}\rangle \delta ^{\pm }A_\alpha ^{{\bf k}%
j}+\sum_\alpha \{\langle \chi _\beta ^{{\bf k}\pm {\bf q}}|\delta ^{\pm
}V_{eff}|\chi _\alpha ^{{\bf k}}\rangle  \nonumber \\
&&+\langle \delta ^{\pm }\chi _\beta ^{{\bf k}\mp {\bf q}}|H-\epsilon _{{\bf %
k}j}|\chi _\alpha ^{{\bf k}}\rangle +\langle \chi _\beta ^{{\bf k}\pm {\bf q}%
}|H-\epsilon _{{\bf k}j}|\delta ^{\pm }\chi _\alpha ^{{\bf k}}\rangle
\}A_\alpha ^{{\bf k}j}=0.  \nonumber \\
&&  \label{2.15}
\end{eqnarray}
This linear system of equations is, in fact, a variation of the original
eigenvalue problem (\ref{2.8}). It determines the position of the minimum of 
$\Lambda $ in the space of the coefficients $\delta ^{\pm }A_\alpha ^{{\bf k}%
j}$, and non of the second--order changes, such as $|\delta ^{\pm }\delta
^{\mp }\chi _\alpha ^{{\bf k}}\rangle $, affect it. The functions $|\delta
^{\pm }\delta ^{\mp }\chi _\alpha ^{{\bf k}}\rangle ,$ on the other hand,
define the value of $\Lambda $ itself in its minimum and must be taken into
account in the evaluation of the dynamical matrix.

We must now solve equation (\ref{2.15}). This equation involves only the 
{\em occupied} states of the unperturbed system, which are necessary for
constructing the induced charge density according to (\ref{2.4}). It may be
solved using an iterative algorithm with the number of operations
proportional to $N_{band}\times N_{basis}^2$, where $N_{band}$ is a number
of filled bands and $N_{basis}$ is a number of the basis functions used for
representing the unperturbed wave functions and their first--order
corrections. This scheme is advantageous when using the LAPW or plane--wave
pseudopotential methods where the conventional matrix--diagonalization
algorithms represent the most time--consuming step which scales as the cube
of the size of the basis. The LMTO\ method, on the other hand, has a small
basis and the inversion of the matrix $\langle \chi _\beta ^{{\bf k}\pm {\bf %
q}}|H-\epsilon _{{\bf k}j}|\chi _\alpha ^{{\bf k}\pm {\bf q}}\rangle $
required for solving Eq. (\ref{2.15}), can easily be performed because its
eigenvalues are $\epsilon _{{\bf k}\pm {\bf q}j^{\prime }}-\epsilon _{{\bf k}%
j}$ and eigenvectors are $A_\alpha ^{{\bf k}\pm {\bf q}j^{\prime }}$, where $%
j^{\prime }=1,N_{basis}$. Because of the minimal size of the basis in the
LMTO\ method, it is not a time--consuming step to find eigenvalues $\epsilon
_{{\bf k}j}$ and eigenvectors $A_\alpha ^{{\bf k}j}$ for all energy bands ($%
=N_{basis}$) at some grid of wave vectors ${\bf k}$. This is independent of
the phonon mode and therefore needs to be done only once. We therefore use
the original eigenstates for the matrix inversion. The result for $\delta
^{\pm }A_\alpha ^{{\bf k}j}$ is then substituted into (\ref{2.11}) which
gives the final expression for $|\delta ^{\pm }{\bf k}j\rangle $ in the form:

\begin{eqnarray}
|\delta ^{\pm }{\bf k}j\rangle  &=&\sum_\alpha |\delta ^{\pm }\chi _\alpha ^{%
{\bf k}}\rangle A_\alpha ^{{\bf k}j}+\sum_{j^{\prime }}\frac{|{\bf k}\pm 
{\bf q}j^{\prime }\rangle }{\epsilon _{{\bf k}j}-\epsilon _{{\bf k}\pm {\bf q%
}j^{\prime }}}\times   \nonumber \\
&&\{\langle {\bf k}\pm {\bf q}j^{\prime }|H-\epsilon _{{\bf k}j}|\sum_\alpha
\delta ^{\pm }\chi _\alpha ^{{\bf k}}A_\alpha ^{{\bf k}j}\rangle +  \nonumber
\\
&&\langle \sum_\alpha \delta ^{\pm }\chi _\alpha ^{{\bf k}\mp {\bf q}%
}A_\alpha ^{{\bf k}\pm {\bf q}j^{\prime }}|H-\epsilon _{{\bf k}j}|{\bf k}%
j\rangle +  \nonumber \\
&&\langle {\bf k}\pm {\bf q}j^{\prime }|\delta ^{\pm }V_{eff}|{\bf k}%
j\rangle \}.  \label{2.16}
\end{eqnarray}
This formula has a simple physical meaning. The first three terms containing 
$|\delta \chi \rangle $ appear because of the use of {\em variational}
solutions. They can be interpreted as incomplete--basis--set corrections to
the last term (the one with $\delta ^{\pm }V_{eff}$), which has the form of
the standard perturbation theory. If all unperturbed states are exact and
they represent {\em mathematically} a complete basis set, then the first and
second terms in (\ref{2.16}) cancel, the third term vanishes and the
standard perturbative formula is recovered. However, if this is not the
case, the use of the functions $|\delta \chi \rangle $ in the basis greatly
reduces the number of states $|{\bf k}\pm {\bf q}j^{\prime }\rangle $ needed
to reach the convergence in (\ref{2.16}). Namely, following the above
derivation, the summation in the last three terms is over $N_{basis}$ energy
states, i.e. over the size of the basis for the unperturbed system$.$ To
illustrate the advantage of this formula we consider the so--called acoustic
sum rule (ASR): suppose all atoms are displaced in the same direction by a
small amount. The change in charge density induced by the rigid movement $%
\nabla V_{eff}$ of the potential will be equal to $\nabla \rho .$ Within
standard perturbation theory one obtains:

\begin{eqnarray}
|\delta {\bf k}j\rangle  &=&\sum_{j^{\prime }}|{\bf k}j^{\prime }\rangle 
\frac{\langle {\bf k}j^{\prime }|\nabla V_{eff}|{\bf k}j\rangle }{\epsilon _{%
{\bf k}j}-\epsilon _{{\bf k}j^{\prime }}}=  \nonumber \\
\  &=&\sum_{j^{\prime }}|{\bf k}j^{\prime }\rangle \langle {\bf k}j^{\prime
}|\nabla |{\bf k}j\rangle =\nabla |{\bf k}j\rangle .  \label{2.12}
\end{eqnarray}
The latter equality can only be obtained if the states $|{\bf k}j^{\prime
}\rangle $ represent a {\em mathematically complete} basis set. This is not
case in the LMTO\ method which employs a minimal basis set to reproduce the
energy bands and wave functions within a certain energy window. On the other
hand, within the minimal\ basis set the ASR can be trivially satisfied if
one uses the expression (\ref{2.16}) for the first-order corrections: here,
by construction $|\delta \chi _\alpha ^{{\bf k}}\rangle \equiv \nabla |\chi
_\alpha ^{{\bf k}}\rangle $ while the last three contributions vanish. (This
is so because they are combined into the integral from a gradient of the
periodic function: $\nabla \{\psi _{{\bf k}j^{\prime }}(H-\epsilon _{{\bf k}%
j})\psi _{{\bf k}j}\}$, which is, by definition, equal to zero.)

The unoccupied states in the expression (\ref{2.16}) should not be
considered as real excitation energies and wave functions. Let us consider
the induced charge density as a ground--state property of both perturbed and
unperturbed systems. In both cases only the occupied states must be well
reproduced, the excited states can, in principle, be arbitrary. The LMTO\
and LAPW\ methods are very suitable for this purpose: they are fast and
accurate within a certain energy window, which is achieved by expanding the
basis functions of the original KKR and augmented--plane--wave (APW)\
methods by Taylor series around some energies $\epsilon _\nu $ at the
centers of interest. {\em The states} $|{\bf k}\pm {\bf q}j^{\prime }\rangle 
$ {\em in} (\ref{2.16}) {\em are the eigenstates of the Hamiltonian matrix} $%
\langle \chi _\beta ^{{\bf k}\pm {\bf q}}|H|\chi _\alpha ^{{\bf k}\pm {\bf q}%
}\rangle $ {\em which is itself constructed to reproduce the occupied energy
bands well}. This is the energy window of interest and all centers of
linearization $\epsilon _\nu $ are in this window. In the KKR and APW
methods the states $|{\bf k}\pm {\bf q}j^{\prime }\rangle $ have the
following meaning: since the KKR (APW) energy bands and eigenvectors are the
eigenstates of the LMTO\ (LAPW) Hamiltonian $\langle \chi _\beta ^{{\bf k}%
}(\epsilon _\nu )|H|\chi _\alpha ^{{\bf k}}(\epsilon _\nu )\rangle $ with $%
\epsilon _\nu \equiv \epsilon _{{\bf k}j}$, the states $|{\bf k}\pm {\bf q}%
j^{\prime }\rangle $ in (\ref{2.16}) will be the eigenstates of the
Hamiltonian $\langle \chi _\beta ^{{\bf k}\pm {\bf q}}(\epsilon _{{\bf k}%
j})|H|\chi _\alpha ^{{\bf k}\pm {\bf q}}(\epsilon _{{\bf k}j})\rangle $ and
only those bands $\epsilon _{{\bf k}\pm {\bf q}j^{\prime }}$ with energy
near $\epsilon _{{\bf k}j}$ will be described correctly. In this case,
finding $|\delta ^{\pm }{\bf k}j\rangle $ requires the knowledge of this
auxiliary spectrum for every occupied energy $\epsilon _{{\bf k}j}.$ We thus
finally conclude that the excited states are not to be interpreted as the
exact ones, {\em only the knowledge of occupied energy bands is necessary}
in our linear--response formulation.

\bigskip\ 

{\bf III. IMPLEMENTATION.}

\smallskip\ 

In this section, an extension of the linear muffin--tin orbital\ method for
linear--response calculations is described. We shall first review the
full--potential LMTO\ method, which is used as the framework in this
implementation. Then, the problem of constructing the changes in the
MT--orbitals due to the atomic movements is discussed. Other problems
considered are the Brillouin--zone integration for metallic systems and the
self--consistency at long wavelengths where the Coulomb singularity $4\pi
/q^2$ makes the standard mixing schemes computationally inefficient.

\smallskip\ 

{\bf a. Full--potential LMTO method. }

\smallskip\ 

We first review the LMTO\ method, which solves the original Schr\"odinger
equation. The space is partitioned into the non overlapping (or slightly
overlapping) muffin--tin spheres $s_R$ surrounding every atom and the
remaining interstitial region $\Omega _{int}.$ Within the spheres, the basis
functions are represented in terms of numerical solutions of the radial
Schr\"odinger equation for the spherical part of the potential multiplied by
spherical harmonics as well as their energy derivatives taken at some set of
energies $\epsilon _\nu $ at the centers of interest. In the interstitial
region, where the potential is essentially flat, the basis functions are
spherical waves taken as the solutions of Helmholtz's equation: $(-\nabla
^2-\epsilon )f(r,\epsilon )=0$ with some fixed value of the average kinetic
energy $\epsilon =\kappa _\nu ^2.$ In particular, in the standard LMTO\
method using the atomic--sphere approximation (ASA)\cite{OKA}, the
approximation $\kappa _\nu ^2=0$ is chosen. In the extensions of the LMTO\
method for a potential of arbitrary shape (full potential), a
multiple--kappa basis set\cite{ManyK} is normally used in order to increase
the variational freedom of the basis functions while recent developments of
a new LMTO technique\cite{NewLMTO} promise to avoid this problem.

The general strategy for including the full--potential terms in the
calculation is the use of the variational principle. A few different
techniques have been developed for taking the non--spherical corrections
into account in the framework of the LMTO method. They include Fourier
transforms of the LMTOs in the interstitial region\cite{Weyrich,Wills},
one--center spherical--harmonics expansions within atomic cells\cite{FORCE},
interpolations in terms of the Hankel functions\cite{Methf} as well as
direct calculations of the charge density in the tight--binding
representation\cite{Bloechl}. In two of these schemes\cite{FORCE,Methf} the
treatment of open structures such as, {\it e.g.} the diamond structure is
complicated and interstitial spheres are usually placed between the atomic
spheres. In the dynamical--matrix calculation it is inconvenient to use
interstitial spheres because they lead to artificial degrees of freedom for
the lattice dynamics. Therefore we will develop the linear--response LMTO\
technique using the plane--wave Fourier representation. This allows us to
apply the method for such materials as $Si$ and $NbC$ without interstitial
spheres$.$ Note, however, that in our previous applications \cite
{PHN,EPI,UFN} for BCC and FCC metals, atomic--cell spherical--harmonic
expansions\cite{FORCE} were used.

Consider the so--called envelope function, which is a singular Hankel
function,

\begin{equation}
K_{\kappa RL}({\bf r}_R-{\bf t})=K_{\kappa Rl}(|{\bf r}_R-{\bf t}|)i^lY_{lm}(%
{\bf r}_R-{\bf t),}  \label{3.1}
\end{equation}
centered at site ${\bf R}+{\bf t}$ and with an energy $\epsilon =\kappa _\nu
^2.$ $Y_{lm}$ denotes a complex spherical harmonic with the phase convention
after Ref. \onlinecite{Varsh}. Inside any other site ${\bf R}^{\prime }+{\bf %
t}^{\prime }$ the Hankel function can be represented as an expansion in
terms of the Bessel functions, $J_{\kappa R^{\prime }L^{\prime }}({\bf r}%
_{R^{\prime }}-{\bf t}^{\prime }),$ i.e

\begin{eqnarray}
K_{\kappa RL}({\bf r}_R-{\bf t}) &=&  \nonumber \\
-\sum_{L^{\prime }}J_{\kappa R^{\prime }L^{\prime }} &&({\bf r}_{R^{\prime
}}-{\bf t}^{\prime })\gamma _{R^{\prime }l^{\prime }}S_{R^{\prime }L^{\prime
}RL}({\bf t}^{\prime }-{\bf t},\kappa ),  \label{3.2}
\end{eqnarray}
where $\gamma _{Rl}=1/s_R(2l+1)$ and $S_{R^{\prime }L^{\prime }RL}({\bf t}%
,\kappa )$ are the structure constants in real space. Note that, while the
index $L\,$ enumerating the basis functions usually runs only over $s,p,$
and $d$ states, the sum over $L^{\prime }$ in this expression must include
higher angular momenta. Normally $l^{\prime }$ goes up to $6-8$. For
convenience, we use the following prefactors in the definitions of the
spherical functions:

\begin{equation}
K_{\kappa Rl}(|{\bf r}_R|)=-\frac{(\kappa s_R)^{l+1}}{(2l-1)!!}h_l(\kappa |%
{\bf r}_R|),  \label{3.3}
\end{equation}

\begin{equation}
J_{\kappa Rl}(|{\bf r}_R|)=\frac{(2l+1)!!}{(\kappa s_R)^l}j_l(\kappa |{\bf r}%
_R|),  \label{3.4}
\end{equation}
where $h_l=j_l-in_l,j_l,$ and $n_l$ are the spherical Hankel, Bessel, and
Neumann functions, respectively. The expression for the structure constants
is then

\begin{eqnarray}
\ &&S_{R^{\prime }L^{\prime }RL}({\bf t},\kappa )=\left( \frac{s_{R^{\prime
}}}w\right) ^{l^{\prime }+1}\left( \frac{s_R}w\right) ^{l+1}\times  \nonumber
\\
&&\ \ \sum_{L^{\prime \prime }}\frac{-4\pi w(2l^{\prime \prime }-1)!!}{%
(2l^{\prime }-1)!!(2l-1)!!}C_{LL^{\prime }}^{L^{\prime \prime }}(\kappa
w)^{l+l^{\prime }-l^{\prime \prime }}\times  \nonumber \\
&&\ \ K_{\kappa wl^{\prime \prime }}(|{\bf t}-{\bf R}^{\prime }+{\bf R}%
|)(-i)^{l^{\prime \prime }}Y_{L^{\prime \prime }}^{*}({\bf t}-{\bf R}%
^{\prime }+{\bf R}),  \label{3.5}
\end{eqnarray}
where $w$ is the average Wigner-Seitz radius and the Hankel function $%
K_{\kappa wl}$ is defined with $w$ instead of $s_R$ in expression (\ref{3.3}%
). The Gaunt coefficients $C_{LL^{\prime }}^{L^{\prime \prime }}$ are
defined by the integral:

\begin{equation}
C_{LL^{\prime }}^{L^{\prime \prime }}=\int Y_LY_{L^{\prime \prime
}}Y_{L^{\prime }}^{*}.  \label{3.6}
\end{equation}

We now consider a Bloch sum of the Hankel functions (\ref{3.1}), centered at
different sites, which, inside the MT-sphere at ${\bf R}^{\prime }$, is
represented by the expansions in the Bessel functions:

\begin{eqnarray}
\sum_t &&e^{i{\bf kt}}K_{\kappa RL}({\bf r}_R-{\bf t})=  \nonumber \\
&&K_{\kappa RL}({\bf r}_R)\delta _{R^{\prime }R}-\sum_{L^{\prime }}J_{\kappa
R^{\prime }L^{\prime }}({\bf r}_{R^{\prime }})\gamma _{R^{\prime }l^{\prime
}}S_{R^{\prime }L^{\prime }RL}^{{\bf k}}(\kappa ),  \label{3.7}
\end{eqnarray}
where $S_{R^{\prime }L^{\prime }RL}^{{\bf k}}(\kappa )$ denotes the lattice
sum of the structure constants (\ref{3.5}). The linear MT-orbitals $|\chi
_{\kappa RL}^{{\bf k}}\rangle $ are now obtained by augmenting the spherical
functions $K_{\kappa RL}$,$J_{\kappa RL}$ in all MT spheres by numerical
radial functions $\Phi _{kRL}^K,\Phi _{kRL}^J$:

\begin{eqnarray}
\chi  &&_{\kappa RL}^{{\bf k}}({\bf r}_{R^{\prime }})=  \nonumber \\
&&\ \Phi _{\kappa RL}^K({\bf r}_R)\delta _{R^{\prime }R}-\sum_{L^{\prime
}}\Phi _{\kappa R^{\prime }L^{\prime }}^J({\bf r}_{R^{\prime }})\gamma
_{R^{\prime }l^{\prime }}S_{R^{\prime }L^{\prime }RL}^{{\bf k}}(\kappa ).
\label{3.8}
\end{eqnarray}
The functions $\Phi _{kRL}^K,\Phi _{kRL}^J$ are the linear combinations of
the solutions $\phi _{RL}({\bf r}_R,\epsilon _{\nu \kappa Rl})\equiv \phi
_{\kappa RL}$ to the radial Schr\"odinger equation as well as their energy
derivatives $\dot \phi _{RL}({\bf r}_R,\epsilon _{\nu \kappa Rl})\equiv \dot 
\phi _{\kappa RL}$ taken at the energies $\epsilon _{\nu \kappa Rl}.$ In the
interstitial region, the linear MT orbitals are represented as multicenter
expansions [left--hand side of Eq. (\ref{3.7})]. In order to calculate the
interstitial--potential matrix elements and represent the charge density, we
use the Fourier transform of the LMTOs in the interstitial region. It is
impossible to consider the Fourier transform of the expression (\ref{3.7})
directly because of the singularities in the Hankel functions. On the other
hand, since this representation will be used for the description of the
basis functions only within $\Omega _{int},$ we can substitute the divergent
part of the Hankel function by a smooth function for $r_R<s_R.$ This regular
function is defined in the Appendix and it is denoted as $\tilde K_{\kappa
RL}.$ We thus introduce a pseudoLMTO $|\tilde \chi _{\kappa RL}^{{\bf k}%
}\rangle $ defined in all space as follows:

\begin{eqnarray}
\tilde \chi _{\kappa RL}^{{\bf k}}({\bf r}) &=&\sum_te^{i{\bf kt}}\tilde K%
_{\kappa L}({\bf r}_R-{\bf t})=  \nonumber \\
\ &=&\sum_G\tilde \chi _{\kappa RL}({\bf k}+{\bf G})e^{i({\bf k}+{\bf G})%
{\bf r}},  \label{3.9}
\end{eqnarray}
which is identical with the true sum (\ref{3.7}) in the interstitial region.

The charge density and the potential have a dual representation:
spherical-harmonic expansions inside the MT-spheres and plane-wave
expansions in the interstitial region. This is usually done by introducing a
smooth pseudocharge density $\tilde \rho $ in all space defined in terms of
the pseudoLMTOs (\ref{3.9}). The pseudodensity coincides with the true
density when ${\bf r}\in \Omega _{int}.\,$In this way, the solution of the
Poisson equation is straightforward and can be done along the lines
developed in Ref. \onlinecite{Weinert}. In practical applications we have
also used the technique described in the Appendix for the Fourier transform
of the Coulomb interactions and for the construction of auxiliary densities.
The exchange--correlation potential is found using the fast Fourier
transform and the interstitial--potential matrix elements are explicitly
evaluated.

\smallskip\ 

{\bf b.} {\bf Changes in the linear muffin-tin orbitals.}

\smallskip\ \ 

We shall now discuss the linear--response calculation. Small displacements
of atoms from their equilibrium positions defined by expression (\ref{2.0})
lead to the change in the Bloch sum of the atom--centered (pseudo) Hankel
functions (\ref{3.9}). Because of the explicit dependence of the basis
functions $|\chi _{\kappa RL}^{{\bf k}}\rangle $ on the atomic positions $%
{\bf R}$, here and in the following the displaced atoms will be denoted by
the index ${\bf \bar R.}$ The change in the Bloch sum of the MT orbitals can
be found analogously to the change in the external potential in Eqs.(\ref
{2.1})--(\ref{2.3}). As a result, we consider two travelling waves with wave
vectors ${\bf k}+{\bf q}$ and ${\bf k}-{\bf q}$, i.e.

\begin{eqnarray}
&&\frac{\delta ^{\pm }\tilde \chi _{\kappa RL}^{{\bf k}}({\bf r})}{\delta 
\bar R_\mu }=-\delta _{\bar RR}\sum_te^{i({\bf k}\pm {\bf q}){\bf t}}\nabla
_\mu \tilde K_{\kappa RL}({\bf r}_R-{\bf t})=  \nonumber \\
&&-\delta _{\bar RR}\sum_Gi(k\pm q+G)_\mu \times \tilde \chi _{\kappa RL}(%
{\bf k}\pm {\bf q}+{\bf G})e^{i({\bf k}\pm {\bf q}+{\bf G}){\bf r}}, 
\nonumber \\
&&  \label{3.10}
\end{eqnarray}
which represent the change of the basis functions in the interstitial region
or the change of the pseudoLMTOs in the whole space. Here we have restored
the original notations: $\delta ^{\pm }\tilde \chi _{\kappa RL}^{{\bf k}}(%
{\bf r})/\delta \bar R_\mu \equiv \delta ^{\pm }\tilde \chi _{\kappa RL}^{%
{\bf k}}({\bf r})$. We also introduce a spherical coordinate system \cite
{Varsh}:

\begin{equation}
{\bf \bar R=}\sum_\mu \bar R_\mu e^\mu =\bar R_{-1}e^{-1}+\bar R_0e^0+\bar R%
_{+1}e^{+1},  \label{3.11}
\end{equation}
which is connected to the Cartesian system as follows: $\bar R_{-1}=+(\bar R%
_x-i\bar R_y)/\sqrt{2},\bar R_0=\bar R_z,\bar R_{+1}=-(\bar R_x+i\bar R_y)/%
\sqrt{2}.$ The reason is that, in the spherical coordinates, the operation $%
\nabla _\mu $ on a product of a radial function $f(r)$ multiplied by the
spherical harmonic takes the simple form:

\begin{eqnarray}
&&\nabla _\mu f(r)Y_{lm}({\bf r})=  \nonumber \\
&&\sqrt{\frac{4\pi }3}C_{lml+1m+\mu }^{1\mu }\left( \frac{df}{dr}-\frac lr%
f(r)\right) Y_{l+1m+\mu }({\bf r})+  \nonumber \\
&&\sqrt{\frac{4\pi }3}C_{lml-1m+\mu }^{1\mu }\left( \frac{df}{dr}+\frac{l+1}r%
f(r)\right) Y_{l-1m+\mu }({\bf r}).  \label{3.12}
\end{eqnarray}

We shall now find variation of the basis functions inside the MT spheres. In
the sphere ${\bf R}^{\prime }$, the original LMTO is defined in the
expression (\ref{3.8}). Its change must include both the changes in the
numerical radial functions and the change in the structure constants:

\begin{eqnarray}
&&\frac{\delta ^{\pm }\chi _{\kappa RL}^{{\bf k}}({\bf r}_{R^{\prime }})}{%
\delta \bar R_\mu }=\frac{\delta ^{\pm }\Phi _{\kappa RL}^K({\bf r}_R)}{%
\delta \bar R_\mu }\delta _{R^{\prime }R}-  \nonumber \\
&&\sum_{L^{\prime }}\frac{\delta ^{\pm }\Phi _{\kappa R^{\prime }L^{\prime
}}^J({\bf r}_{R^{\prime }})}{\delta \bar R_\mu }\gamma _{R^{\prime
}l^{\prime }}S_{R^{\prime }L^{\prime }RL}^{{\bf k}}(\kappa )-  \nonumber \\
&&\sum_{L^{\prime }}\Phi _{\kappa R^{\prime }L^{\prime }}^J({\bf r}%
_{R^{\prime }})\gamma _{R^{\prime }l^{\prime }}\frac{\delta ^{\pm
}S_{R^{\prime }L^{\prime }RL}^{{\bf k}}(\kappa )}{\delta \bar R_\mu }.
\label{3.13}
\end{eqnarray}
The change in the numerical functions contains two contributions. Since $%
\Phi _{\kappa RL}^K$, $\Phi _{\kappa RL}^J$ are constructed from the
solutions of the radial Schr\"odinger equation and their energy derivatives, 
$\phi _{\kappa RL}$ and $\dot \phi _{\kappa RL},$ the change in $\phi
_{\kappa RL}$ and $\dot \phi _{\kappa RL}$ is a result of both the rigid
movement of the spherical component of the potential and the change in the
shape of the spherical component. In the following, it is convenient to
treat the rigid movements of the potential within the MT--sphere centered at 
${\bf R}$ separately, i.e. represent the total change in the form:

\begin{equation}
\frac{\delta ^{\pm }V_{eff}({\bf r}_R)}{\delta \bar R_\mu }=-\delta _{\bar R%
R}\nabla _\mu V_{eff}({\bf r}_R)+\frac{\delta _{(s)}^{\pm }V_{eff}({\bf r}_R)%
}{\delta \bar R_\mu },  \label{3.14}
\end{equation}
where the notation $\delta _{(s)}^{\pm }$ stands for the ''soft'' change,
i.e. the variation connected with the change in the shape of the function.
The functions $\delta ^{\pm }\phi _{\kappa RL}/\delta \bar R_\mu $ are
represented in a form similar to (\ref{3.14}), i.e. $\delta ^{\pm }\phi
_{\kappa RL}/\delta \bar R_\mu =-\delta _{\bar RR}\nabla _\mu \phi _{\kappa
RL}+\delta _{(s)}^{\pm }\phi _{\kappa RL}/\delta \bar R_\mu $, where the
last (soft) contribution is found by solving the radial Sternheimer equation:

\begin{eqnarray}
&&(-\nabla _r^2+\frac{l(l+1)}{r^2}+V_{eff}^{SPH}-\epsilon _{\nu \kappa Rl})%
\frac{\delta _{(s)}^{\pm }\phi _{\kappa RL}}{\delta \bar R_\mu }+  \nonumber
\\
&&\left( \frac{\delta _{(s)}^{\pm }V_{eff}^{SPH}}{\delta \bar R_\mu }-\frac{%
\delta _{(s)}^{\pm }\epsilon _{\nu \kappa Rl}}{\delta \bar R_\mu }\right)
\phi _{\kappa RL}=0.  \label{3.15}
\end{eqnarray}
The superscript ''$SPH$'' here denotes the spherical component of the
potential and the perturbation. It is, in principle, not a problem to take
all non--spherical terms of the perturbation into account. If this is done,
the first--order changes in the radial functions are no longer given by a
single spherical harmonic but as an expansion in $Y_{lm}$. One obtains an 
{\em uncoupled }system of radial equations, which can easily be solved \cite
{SSC}. However, in the problem of lattice dynamics the change in $\phi
_{\kappa RL}$ and $\dot \phi _{\kappa RL}$\thinspace due to the change in
the shape of the spherical component of the potential is small. This is
because the motions of atoms mainly distort the dipole part of the
potential. If the change in the shape of the spherical component can be
described by some constant shift of the energy, it may be cancelled by
appropriate choice of the change $\delta _{(s)}^{\pm }\epsilon _{\nu \kappa
Rl}/\delta \bar R_\mu $ in the energies $\epsilon _{\nu \kappa Rl}.$ This
cancellation can, for instance, be obtained by finding $\delta _{(s)}^{\pm
}\epsilon _{\nu \kappa Rl}/\delta \bar R_\mu $ with fixed logarithmic
derivatives $D_{\nu \kappa Rl}$\thinspace .\thinspace (The derivatives $%
D_{\nu \kappa Rl}$ are evaluated at the occupied centers of gravities of the
bands for the unperturbed crystal.) We thus see that the influence of the
constant shifts to the change in the basis set can be eliminated and,
therefore, one can neglect by the contribution $\delta _{(s)}^{\pm }\phi
_{\kappa RL}/\delta \bar R_\mu $ in practical calculations. The accuracy of
this approximation is quite good which has already been confirmed by good
agreement between total energy and force calculations with the original
LMTO\ method\cite{FORCE} where the same approximation was used for deriving
the force formula.

We now give the formula for the change in the structure constants which
enters Eq. (\ref{3.13}). It is expressed as the difference between the
gradients of the structure constants for wave vectors ${\bf k}$ and ${\bf k}%
\pm {\bf q,}$ i.e

\begin{eqnarray}
&&\frac{\delta ^{\pm }S_{R^{\prime }L^{\prime }RL}^{{\bf k}}(\kappa )}{%
\delta \bar R_\mu }=  \nonumber \\
&&\delta _{\bar RR^{\prime }}\nabla _\mu S_{R^{\prime }L^{\prime }RL}^{{\bf k%
}}(\kappa )-\delta _{\bar RR}\nabla _\mu S_{R^{\prime }L^{\prime }RL}^{{\bf k%
}\pm {\bf q}}(\kappa ).  \label{3.16}
\end{eqnarray}
The gradient is with respect to ${\bf R}^{\prime }-{\bf R}$. From (\ref{3.12}%
), using the recursion relations for the Hankel functions, it follows that
the change in the structure constants can be expressed in terms of the
structure constants:

\begin{eqnarray}
&&\nabla _\mu S_{R^{\prime }l^{\prime }m^{\prime }Rlm}^{{\bf k}}(\kappa )= 
\nonumber \\
\ =i &&\sqrt{\frac{4\pi }3}C_{l^{\prime }-1m^{\prime }-\mu l^{\prime
}m^{\prime }}^{1\mu }\frac{\kappa ^2s_{R^{\prime }}}{(2l^{\prime }-1)}%
S_{R^{\prime }l^{\prime }-1m^{\prime }-\mu Rlm}^{{\bf k}}(\kappa )+ 
\nonumber \\
+i &&\sqrt{\frac{4\pi }3}C_{l^{\prime }+1m^{\prime }-\mu l^{\prime
}m^{\prime }}^{1\mu }\frac{(2l^{\prime }+1)}{s_{R^{\prime }}}S_{R^{\prime
}l^{\prime }+1m^{\prime }-\mu Rlm}^{{\bf k}}(\kappa ).  \label{3.17}
\end{eqnarray}
Here, the left index of the structure constants has changed to $l^{\prime
}\pm 1,m^{\prime }-\mu $, but the right index $lm\,$remains the same. An
analogous formula exist in which the right index change is $l\pm 1,m+\mu $
in and the left index is unchanged.

It is seen that the change in the MT orbital (\ref{3.13}) can be represented
as a rigid part, a small soft part and a contribution from the change in the
structure constants:

\begin{eqnarray}
\frac{\delta ^{\pm }\chi _{\kappa RL}^{{\bf k}}({\bf r}_{R^{\prime }})}{%
\delta \bar R_\mu } &=&-\delta _{\bar RR^{\prime }}\nabla _\mu \chi _{\kappa
RL}^{{\bf k}}({\bf r}_{R^{\prime }})+\frac{\delta _{(s)}^{\pm }\chi _{\kappa
RL}^{{\bf k}}({\bf r}_{R^{\prime }})}{\delta \bar R_\mu }-  \nonumber \\
&&\sum_{L^{\prime }}\Phi _{\kappa R^{\prime }L^{\prime }}^J({\bf r}%
_{R^{\prime }})\gamma _{R^{\prime }l^{\prime }}\frac{\delta ^{\pm
}S_{R^{\prime }L^{\prime }RL}^{{\bf k}}(\kappa )}{\delta \bar R_\mu }.
\label{3.18}
\end{eqnarray}
It is convenient to separate the rigid part since it gives rise to a rigid
contribution in the electronic response:

\begin{equation}
\frac{\delta ^{\pm }\rho ({\bf r}_R)}{\delta \bar R_\mu }=-\delta _{\bar R%
R}\nabla _\mu \rho ({\bf r}_R)+\frac{\delta _{(s)}^{\pm }\rho ({\bf r}_R)}{%
\delta \bar R_\mu }.  \label{3.19}
\end{equation}
Since the induced charge density (\ref{3.19}) has the same form as the
change in the potential (\ref{3.14}), we need not calculate the gradients of
the charge density and the potential. This is important since these
gradients are huge in the core region, which could result in large numerical
errors. The second term in (\ref{3.18}) is $\delta _{(s)}^{\pm }\chi
_{\kappa RL}^{{\bf k}}({\bf r}_{R^{\prime }})/\delta \bar R_\mu $. It is
constructed from the changes $\delta _{(s)}^{\pm }\phi _{\kappa RL}/\delta 
\bar R_\mu $ and their energy derivatives which are numerically small. This
function is exactly equal to zero together with its first-order radial
derivative at the sphere $s_{R^{\prime }}$. It translates like a Bloch wave
with vector ${\bf k}\pm {\bf q}$ because the original form of one--center
expansion (\ref{3.8}) translates with wave vector ${\bf k}$ while the
first--order changes $\delta _{(s)}^{\pm }\phi _{\kappa RL}/\delta \bar R%
_\mu $ translate, like the perturbation, with wave vector $\pm {\bf q}$. The
whole expression (\ref{3.18}) also translates with wave vector ${\bf k}\pm 
{\bf q}$ and fits into the multicenter expansion of the change in the basis
set in the interstitial region [formula (\ref{3.10})]. However, since the
original LMTOs are continuous and only differentiable to first order at the
boundaries of the MT spheres, the matching of the change in the basis set is
only continuous but not differentiable. This, in principle, leads to a kink
in the change of the charge density. However, it does not have any effect in
the calculation of the dynamical matrix if the latter is compared with the
second--order derivative of the total energy derived from the frozen-phonon
supercell calculation. This is so because the extension of the LMTO\ method
described here is just an analytical version of the finite-difference
approach employed in the supercell technique. When applied to the same
problem, the results of both approaches have to be the same except for the
errors introduced by taking finite differences. This concerns the comparison
of not only the dynamical matrix and the phonon frequencies, but also the
changes in the basis set, the expansion coefficients, the charge densities
and in all other quantities which can be obtained by the frozen--phonon
LMTO\ technique.

We now turn to the problem of calculating the change in the expansion
coefficients $A_{\kappa RL}^{{\bf k}j},$ which are necessary to compute the
first--order corrections according to (\ref{2.11}). From expression (\ref
{2.16}), the change $\delta ^{\pm }A_{\kappa RL}^{{\bf k}j}/\delta \bar R%
_\mu $ is given by

\begin{eqnarray}
\frac{\delta ^{\pm }A_{\kappa RL}^{{\bf k}j}}{\delta \bar R_\mu }
&=&\sum_{j^{\prime }}\frac{A_{\kappa RL}^{{\bf k}\pm {\bf q}j^{\prime }}}{%
\epsilon _{{\bf k}j}-\epsilon _{{\bf k}\pm {\bf q}j^{\prime }}}  \nonumber \\
&&\times \left( \frac{\delta ^{\pm }H^{{\bf k}\pm {\bf q}j^{\prime }{\bf k}j}%
}{\delta \bar R_\mu }-\epsilon _{{\bf k}j}\frac{\delta ^{\pm }O^{{\bf k}\pm 
{\bf q}j^{\prime }{\bf k}j}}{\delta \bar R_\mu }\right)   \label{3.20}
\end{eqnarray}
and it is expressed in terms of the change in the hamiltonian and the
overlap matrices. Here, the change in the matrix elements of the hamiltonian
is given by the band representation:

\begin{equation}
\frac{\delta ^{\pm }H^{{\bf k}\pm {\bf q}j^{\prime }{\bf k}j}}{\delta \bar R%
_\mu }=\sum_{\kappa ^{\prime }R^{\prime }L^{\prime }}\sum_{\kappa
RL}A_{\kappa ^{\prime }R^{\prime }L^{\prime }}^{{\bf k}\pm {\bf q}j^{\prime
}*}\frac{\delta ^{\pm }H_{\kappa ^{\prime }R^{\prime }L^{\prime }\kappa RL}}{%
\delta \bar R_\mu }A_{\kappa RL}^{{\bf k}j}  \label{3.21}
\end{equation}
and a similar formula holds for the matrix elements of the overlap integral.
In the original, $\{\kappa RL\},$ representation the changes $\delta ^{\pm
}H_{\kappa ^{\prime }R^{\prime }L^{\prime }\kappa RL}/\delta \bar R_\mu $ and%
$\,\delta ^{\pm }O_{\kappa ^{\prime }R^{\prime }L^{\prime }\kappa RL}/\delta 
\bar R_\mu $ are readily computed using the formulae (\ref{3.10}) and (\ref
{3.18}) for the first--order changes in the basis set. It is indeed even
more advantageous to find the corresponding formulae by directly varying the
expressions for the hamiltonian and the overlap matrices. This avoids the
problem of combining the contributions with the gradients of numerical
radial functions to the surface integrals. One point about calculating the
change in the interstitial kinetic--energy matrix elements and the
interstitial overlap integrals is worth noticing. Since these matrix
elements contain energy--derivative of the structure constants, the change
in these matrix elements will contain the change in this energy derivative.
The corresponding formula can be found by taking the derivative with respect
to $\kappa ^2$ in the expressions (\ref{3.16}) and (\ref{3.17}).

Another problem is to find second--order changes in the LMTO\ basis
functions as well as second--order variations in the Hamiltonian and the
overlap matrices. They are necessary for computing the
incomplete--basis--set corrections in (\ref{2.9}) for the dynamical matrix.
In the interstitial region the second--order change in the pseudoLMTOs is
simply given by:

\begin{eqnarray}
\frac{\delta ^{+}\delta ^{-}\chi _{\kappa RL}^{{\bf k}}({\bf r})}{\delta 
\bar R_{\mu ^{\prime }}^{\prime }\delta \bar R_\mu } &=&\delta _{\bar R%
^{\prime }R}\delta _{\bar RR}\sum_Gi(k+G)_{\mu ^{\prime }}i(k+G)_\mu 
\nonumber \\
&&\times \chi _{\kappa RL}({\bf k}+{\bf G})e^{i({\bf k}+{\bf G}){\bf r}}.
\label{3.A.1}
\end{eqnarray}
Inside the MT spheres the expression is more complicated, but can be found
straightforwardly by performing one more variation $\delta ^{-}/\delta \bar R%
_{\mu ^{\prime }}$ of expression (\ref{3.13}) for the first--order change.
It will contain second--order changes in the numerical radial functions and
second--order changes in the structure constants as well as different
products of the first--order changes in these quantities. The second--order
changes in the structure constants are given by

\begin{eqnarray}
&&\frac{\delta ^{+}\delta ^{-}S_{R^{\prime }L^{\prime }RL}^{{\bf k}}(\kappa )%
}{\delta \bar R_{\mu ^{\prime }}^{\prime }\delta \bar R_\mu }=  \nonumber \\
&&\ \delta _{\bar RR^{\prime }}\nabla _\mu \frac{\delta ^{-}S_{R^{\prime
}L^{\prime }RL}^{{\bf k}}(\kappa )}{\delta \bar R_{\mu ^{\prime }}^{\prime }}%
-\delta _{\bar RR}\nabla _\mu \frac{\delta ^{-}S_{R^{\prime }L^{\prime }RL}^{%
{\bf k}+{\bf q}}(\kappa )}{\delta \bar R_{\mu ^{\prime }}^{\prime }}.
\label{3.A.2}
\end{eqnarray}
This is obtained from the expression (\ref{3.16}) and $\delta ^{-}S^{{\bf k}}
$ is expressed via the difference between the gradients of the structure
constants for the wave vectors ${\bf k}\,$ and ${\bf k}-{\bf q}$, while $%
\delta ^{-}S^{{\bf k}+{\bf q}}$ is the difference between $\nabla S\,$ for
the wave vectors ${\bf k}+{\bf q}\,$ and ${\bf k.}$ Alternatively, the
expression (\ref{3.A.2}) can be found by first considering the expression
for the structure constants in the supercell and then, assuming the form (%
\ref{2.0}) for the atomic displacements, transferring the supercell
expression to the original structure. The second--order gradients $\nabla
_{\mu ^{\prime }}\nabla _\mu \,S$ are calculated using (\ref{3.17}) and they
are again the structure constants with the left index changed to $l^{\prime
}\pm 2,m^{\prime }-\mu ^{\prime }-\mu $ and unchanged right index.
Analogously, they can be expressed in terms of the structure constants of
the same left index $l^{\prime }m^{\prime }$ and the right index: $l\pm
2,m+\mu ^{\prime }+\mu .$

The second--order changes in the numerical radial functions must also be
calculated. They contain contributions $\nabla _{\mu ^{\prime }}\nabla _\mu
\Phi _{\kappa RL}^{K,J}$ due to the rigid movement of the spherical part of
the potential to second order, changes due to the rigid movements of the
first--order variations in the shape of the spherical part (rigid movement
of the soft part), as well as the contributions arising from the change in
the shape of the spherical part to second order (second--order soft part).
As we discussed above, one can neglect by the influence of the change in the
shape of $V_{eff}^{SPH}$ to the change in the basis. Therefore, we must only
keep the rigid contributions described by $\nabla _{\mu ^{\prime }}\nabla
_\mu \Phi _{\kappa RL}^{K,J}.$

\smallskip\ \ 

{\bf c. Brillouin-zone integrals.}

\smallskip\ 

After computing the first--order corrections to the wave functions, we have
to perform the ${\bf k}$-space integration over the first Brillouin zone
(BZ) in order to find the change in the charge density from Eq. (\ref{2.5}).
The BZ integration is also required for calculating the incomplete-basis-set
corrections to the dynamical matrix. It is in general a full--zone
integration while for the high--symmetry wave vectors the integrals are
reduced to that portion of the BZ which is irreducible with respect to the
symmetry of the perturbation vector.

Two kinds of the integrals have to be performed in the linear--response
calculation. The first one has the following form

\begin{equation}
I_1({\bf q})=\sum_{{\bf k}j}2f_{{\bf k}j}A_{kj}({\bf q})  \label{3.22}
\end{equation}
and the second one is given by

\begin{equation}
I_2({\bf q})=\sum_{{\bf k}jj^{\prime }}\frac{2f_{{\bf k}j}(1-f_{{\bf k}\pm 
{\bf q}j^{\prime }})}{\epsilon _{{\bf k}j}-\epsilon _{{\bf k}\pm {\bf q}%
j^{\prime }}}M^{{\bf k}\pm {\bf q}j^{\prime }{\bf k}j}.  \label{3.23}
\end{equation}
where $A_{kj}({\bf q})$ and $M^{{\bf k}\pm {\bf q}j^{\prime }{\bf k}j}$ are
the matrix elements which presumably are smooth functions of wave vectors.
In order to calculate these integrals we use the tetrahedron method in Ref. %
\onlinecite{TETR1}. In this method, the BZ is set up by the
reciprocal--lattice translational vectors and it is divided into small
primitive cells exactly as in standard fast--Fourier--transform analysis.
The calculation becomes simpler if the ${\bf q}$ vector coincides with a
mesh point because ${\bf k}\pm {\bf q}$ vectors are also mesh points. In
this way the energy bands, the expansion coefficients, and the structure
constants have to be calculated only once at the mesh of the irreducible
wave vectors ${\bf k}$ for the unperturbed crystal. Applying symmetry
operations, these quantities can be found for any general ${\bf k}$.

When applied to a semiconductor, the tetrahedron method is identical to the
special--point method of Monkhorst and Pack\cite{Monk}, which means that the
occupation numbers $f_{{\bf k}j}$ in (\ref{3.22}) and (\ref{3.23}) can be
regarded as the geometrical weights of the ${\bf k}$--points. Both integrals
(\ref{3.22}) and (\ref{3.23}) converge rapidly with respect to the number of 
${\bf k}$--points. The integral $I_2({\bf q})$ reduces to the integral $I_1(%
{\bf q})$ by performing the summation over the unoccupied bands $j^{\prime }$%
.

For metallic systems a significantly larger number of ${\bf k}$-points $%
(N_k) $ is necessary when the matrix elements as well as the energy
denominator $\epsilon _{{\bf k}j}-\epsilon _{{\bf k}\pm {\bf q}j^{\prime }}$
are interpolated linearly within the tetrahedron. For these systems there
are two sources of errors: the first is connected with the interpolation of
the matrix elements and the second is connected with the interpolation of
the Fermi surface. The latter can easily be circumvented in the
linear--response calculation, since the Fermi surface can be determined
accurately from the band structure of the unperturbed crystal. For the
integrals $I_1({\bf q})$ this can be done using the method described in Ref. %
\onlinecite{TETR2} which is based on considering two, coarse and dense,
meshes. In the tetrahedron method the integration weight of a particular $%
{\bf k}$--point is calculated by integrating over the occupied parts of
those tetrahedra that contain this point. The occupied part of the
tetrahedron is found by linear interpolation between the energies at the
corners of this tetrahedron. Suppose we introduce a much denser mesh that
also contains the original coarse mesh. We will need only the energies $%
\epsilon _{{\bf k}j}$ at this dense mesh, which will define the accurate
Fermi surface (for example also by linear interpolation). Then, the occupied
part of the tetrahedron at the coarse grid can be found by not interpolating
linearly the energies known at its corners but as a piece of the accurate
Fermi surface found with help of the dense grid. The same is applicable to
the integrals $I_2({\bf q}):$ we consider the dense and the coarse grids.
The band energies are known at the dense grid. To find the integration
weights we must find a region in the tetrahedron where the state $|{\bf k}%
j\rangle $ is occupied and the state $|{\bf k}\pm {\bf q}j\rangle $ is
unoccupied. This can be done using the linear interpolation but on the dense
grid. We must also include the energy denominator $\epsilon _{{\bf k}%
j}-\epsilon _{{\bf k}\pm {\bf q}j^{\prime }}$. This is also interpolated
linearly but again using the dense grid. Consequently, all the effects from
the energy bands and the Fermi surface are treated exactly in such scheme
which allows us to avoid this source of errors in the integration.

Another source of errors is connected with the linear interpolation of the
matrix elements. We have already mentioned that the matrix elements are
normally smooth functions of wave vectors and one can expect that after
eliminating the errors connected with the approximate treatment of the Fermi
surface, the number of ${\bf k}$--points need not be too large. However, in
practical calculations a large cancellation occurs between the two kinds of
the integrals, (\ref{3.22}) and (\ref{3.23}). If one uses different
integration weights, it will lead to large numerical errors connected with
the different convergency with respect to $N_k$ in these integrals. Our task
is thus to extract a large contribution from the integral of the type $I_2(%
{\bf q})$ and reduce it to the form $I_1({\bf q}).$ This is achieved by
rewriting the energy denominator $\Delta =\epsilon _{{\bf k}j}-\epsilon _{%
{\bf k}\pm {\bf q}j^{\prime }}$ in the expression (\ref{3.23}) as follows:

\begin{equation}
\frac 1\Delta =\frac \Delta {\Delta ^2+\delta ^2}+\frac 1\Delta \left( 1-%
\frac{\Delta ^2}{\Delta ^2+\delta ^2}\right) ,  \label{3.24}
\end{equation}
where the broadening $\delta $ is usually chosen $\sim 0.1Ry.$ Then, the sum
over unoccupied bands $j^{\prime }$ in the integral containing $\Delta
/(\Delta ^2+\delta ^2)$ is readily performed because this expression remains
regular when $\Delta \rightarrow 0.$ Consequently, this integral is reduced
to the integral of the type $I_1({\bf q}).$ The second integral in (\ref
{3.24}) contains $1/\Delta $ and must be treated as the integral of the type 
$I_2({\bf q})$ where the original matrix element $M^{{\bf k}\pm {\bf q}%
j^{\prime }{\bf k}j}$ is now multiplied by the expression in brackets in (%
\ref{3.24}). However, because the latter rapidly goes to zero for $\Delta
\gg \delta $, the whole integral remains small and it is non zero only for
the band transitions $j\rightarrow j^{\prime }$ between the states near the
Fermi level. In practical calculations of the dynamical matrix, this
procedure allows us to avoid the errors connecting with the large
cancellations.

We finally mention that a simple correction formula which significantly
improves the convergency of the integrals $I_1({\bf q})$ by taking into
account the curvature of the matrix elements beyond the linear interpolation
was derived by Bl\"ochl\cite{Bloechl,TETR2}$.$ Unfortunately, it is hard to
derive such a correction for the integrals $I_2({\bf q})$ because of the
appearance of the energy denominator but we always use the Bl\"ochl
correction for the integrals (\ref{3.22}).

\smallskip\ 

{\bf d. Self-consistency at long wavelengths.}

\smallskip\ 

The change in the charge density (\ref{2.4}) induced by the displacements of
nuclei screens the external perturbation (\ref{2.3}), and the
linear--response equations (\ref{2.5})--(\ref{2.7}) must, therefore, be
solved self--consistently. Let us assume that we have found the response of
the electrons, $\delta \rho ^{(0)},$ to the external perturbation $\delta
V_{ext\text{ }}$or the perturbation screened by some guessed $\delta \rho
^{guess}$ (here we omit ''$\pm $'' for simplicity). The latter could, for
instance, be the rigid shifts of the charge density around the displaced
nuclei and in practical calculations the external perturbation is always
considered as the change in the bare Coulomb potential (\ref{2.3}) plus the
term $\nabla \rho $ within the MT sphere. The response $\delta \rho ^{(0)}$
is found along the lines described above and, consequently, it can be
considered as some polarization operator $\hat \Pi $ that acts on $\delta
V_{ext\text{ }},\,$i.e.

\begin{equation}
\delta \rho ^{(0)}=\hat \Pi \delta V_{ext\text{ }}.  \label{3.25}
\end{equation}
If we omit the terms containing the change in the basis functions and forget
about the completeness problem of the unperturbed states, the operator $\hat 
\Pi $ is given by the independent--particle polarizability function $\hat \pi
$:

\begin{eqnarray}
\pi _{{\bf q}}({\bf r},{\bf r}^{\prime }) &=&\sum_{{\bf k}jj^{\prime }}\frac{%
f_{{\bf k}j}-f_{{\bf k}+{\bf q}j^{\prime }}}{\epsilon _{{\bf k}j}-\epsilon _{%
{\bf k}+{\bf q}j^{\prime }}}  \nonumber \\
&&\times \psi _{{\bf k}+{\bf q}j^{\prime }}({\bf r})\psi _{{\bf k}j}^{*}(%
{\bf r})\psi _{{\bf k}+{\bf q}j^{\prime }}^{*}({\bf r}^{\prime })\psi _{{\bf %
k}j}({\bf r}^{\prime }).  \label{3.26}
\end{eqnarray}
The operator $\pi $ is an integral operator while $\hat \Pi \,$ is not
necessarily one. It denotes the procedure how to construct the change $%
\delta \rho $ from $\delta V_{ext\text{ }}.$ In particular, $\hat \Pi $
contains those part of the operator $\pi $ in which the sum over conduction
states runs only over the number which is equal to the number of the basis
functions, $N_{basis}.$ It also contains the contribution from the change in
the basis functions according to (\ref{3.10}), (\ref{3.18}).

After the initial response $\delta \rho ^{(0)}$ has been found, we have to
calculate the screened perturbation (\ref{2.6}). Let us call the Coulomb
interaction, $e^2/|{\bf r}-{\bf r}^{\prime }|,$ for $v_C$ and the
exchange--correlation interaction in the LDA, $dV_{xc}/d\rho \times \delta (%
{\bf r}-{\bf r}^{\prime }),$ for $v_{xc}.$ Then, the change $\delta V_{eff}$
can be written as follows:

\begin{equation}
\delta V_{eff}=\delta V_{ext}+(v_C+v_{xc})\delta \rho ,  \label{3.27}
\end{equation}
and the new electronic response $\delta \rho =\hat \Pi \delta V_{eff}.\,$ We
thus see that the self-consistency of the induced charge density means
solving the Dyson equation:

\begin{equation}
\delta \rho =\delta \rho ^{(0)}+\hat \Pi (v_C+v_{xc})\delta \rho .
\label{3.28}
\end{equation}
When $q\rightarrow 0,$ the integral $v_C\delta \rho $ diverges as $1/q^2$
which immediately means that searching for the solution of Eq. (\ref{3.28})
by iterations, i.e. $\delta \rho =\delta \rho ^{(0)}+\hat \Pi
(v_C+v_{xc})\delta \rho ^{(0)}+...,$ is impossible. However, it is possible
when the input to the next, ($i+1)$th$,$ iteration is prepared by mixing the
output and input densities from the previous, $(i)$th, iteration, i.e. $%
\delta \rho _{i+1}^{inp}=\lambda _{mix}\delta \rho _i^{out}+(1-\lambda
_{mix})\delta \rho _i^{inp},$ but the mixing parameter $\lambda _{mix}$ must
be chosen to be proportional to $q^2.$ This makes the standard mixing
schemes in the long-wavelength limit extremely time--consuming.

This divergency problem is well known, and in the dielectric--matrix
approach it is avoided by writing the solution (\ref{3.28}) in the form:

\begin{equation}
\delta \rho =\epsilon ^{-1}\delta \rho ^{(0)},  \label{3.29}
\end{equation}
where $\epsilon ^{-1}=(1-\Pi v_C-\Pi v_{xc})^{-1}$ is an inverse dielectric
matrix of the crystal. [The relation (\ref{3.29}) is usually written for the
potentials $\delta V_{eff}$ and $\delta V_{ext}$ but in the present context
it is more convenient to remain within the density language.] If for a metal 
$\Pi v_C$ is proportional to $N(\epsilon _F)/q^2$, where $N(\epsilon _F)$ is
the density of states at the Fermi energy $\epsilon _F$, then $\epsilon ^{-1}
$ behaves as $q^2$ when $q\rightarrow 0\,$: this is the well--known
long-wavelength behavior of the metallic dielectric function. What we
actually do when solving (\ref{3.28}) by iterations is trying to sum up $%
1-x+x^2-...=1/(1+x)$ for $x>1.$

In order to avoid this problem we use Thomas--Fermi--like screening theory.
To explain the idea we assume that the change in charge density and the
potential are expanded in plane waves,

\begin{equation}
\delta \rho ({\bf r})=\sum_G\delta \rho ({\bf G})e^{i({\bf q}+{\bf G}){\bf r}%
}.  \label{3.30}
\end{equation}
We divide the Coulomb interaction $v_C$ into a long-range and a short-range
parts, i.e $v_C=v_C^{long}+v_C^{short},$ where $v_C^{long}=4\pi
e^2/q^2\times exp[i{\bf q}({\bf r}-{\bf r}^{\prime })].$ The
exchange-correlation in LDA is always short ranged and can be treated
together with $v_C^{short}$, i.e. $v_C^{short}+v_{xc}=w^{short}.$ The Dyson
equation can then be written as follows:

\begin{equation}
\delta \rho ({\bf r})=\delta \rho ^{(0)}({\bf r})+\frac{4\pi e^2}{q^2}\delta
\rho (0)\Pi _{{\bf q}}({\bf r})+\{\hat \Pi w^{short}\delta \rho \}({\bf r}),
\label{3.31}
\end{equation}
where we have separated out the divergent contribution, $\delta \rho
(0)\equiv \delta \rho ({\bf G}=0),$ and where we have called the response of
electrons to the perturbation given by a single plane wave $exp(i{\bf qr})$
for $\Pi _{{\bf q}}({\bf r}).$ It can be written as an integral over the
unit cell $\Omega _c$:

\begin{equation}
\Pi _{{\bf q}}({\bf r)=}\int_{\Omega _c}\pi _{{\bf q}}({\bf r},{\bf r}%
^{\prime })e^{i{\bf qr}^{\prime }}d{\bf r}^{\prime }.  \label{3.32}
\end{equation}
The ${\bf G}=0$ part of Eq. (\ref{3.31}) can be written as follows:

\begin{equation}
\delta \rho (0)=\epsilon _{long}^{-1}\left( \delta \rho ^{(0)}(0)+\{\hat \Pi
w^{short}\delta \rho \}({\bf G}=0)\right) ,  \label{3.33}
\end{equation}
where we have introduced an effective dielectric constant:

\begin{equation}
\epsilon _{long}=1-\frac{4\pi e^2}{q^2}\Pi _{{\bf q}}({\bf G}=0).
\label{3.34}
\end{equation}
Inserting Eq. (\ref{3.33}) in to the Dyson equation (\ref{3.31}) we obtain:

\begin{eqnarray}
&&\delta \rho ({\bf r})=\delta \rho ^{(0)}({\bf r})+\frac{4\pi e^2}{%
q^2+\kappa _D^2}  \nonumber \\
&&\times \left( \delta \rho ^{(0)}(0)+\{\hat \Pi w^{short}\delta \rho \}(%
{\bf G}=0)\right)   \nonumber \\
&&\times \Pi _{{\bf q}}({\bf r})+\{\hat \Pi w^{short}\delta \rho \}({\bf r}),
\label{3.35}
\end{eqnarray}
where $\kappa _D^2=-4\pi e^2\Pi _{{\bf q}}({\bf G}=0)$ is the Debye
screening radius$.$ The screened Dyson equation (\ref{3.35}) is free of the
difficulties discussed above and can be solved iteratively. First, one has
to find the function $\Pi _{{\bf q}}({\bf r})$ as the response of electrons
to a single plane wave $exp(i{\bf qr}),$ and from that obtain $\kappa _D^2.$
Then the initial distribution $\delta \rho ^{(0)}({\bf r})$ is calculated$.$
During the iterations one first finds the response to the short--range part
of the perturbation, i.e. $\{\hat \Pi w^{short}\delta \rho \}({\bf r})$,
and, secondly, the long-wavelength contribution is added as given by the
second term in the right--hand side of Eq. (\ref{3.35}). The output change
in the charge density is usually mixed with the input $\delta \rho $ to
obtain an input for the new iteration. This makes the self--consistent cycle
stable, but the mixing parameter $\lambda _{mix}\,$in this case does not
have to go to zero for ${\bf q}\rightarrow 0\,$ and it is usually chosen to
be $0.2-0.5.$ In practical applications we have found that the number of
iterations required to solve (\ref{3.35}) is about $10$ while for solving
the original Dyson equation (\ref{3.28}) the number of iterations varies
from $50$ to $200$ depending on the length $|{\bf q|}$ of the wave vector .
The latter is, of course, not true for those phonon modes where $\delta \rho
(0)\equiv 0$ by symmetry$.$

One can obviously consider the screening of not only the component $\delta V(%
{\bf G})\sim \delta \rho ({\bf G})/|{\bf q}+{\bf G}|^2$ with ${\bf G}=0$ but
all the components within a certain sphere $|{\bf q}+{\bf G}|\leq G_{cutoff}.
$ This, for instance, is necessary for those zone--boundary wave vectors
where $|{\bf q}|=|{\bf q}+{\bf G}|.\,$In this case the function $\Pi _{{\bf q%
}}({\bf r})$ is replaced by the functions $\Pi _{{\bf q}+{\bf G}}({\bf r})$,
i.e. at the beginning it is necessary to calculate the response of the
electrons to the perturbation $exp[i({\bf q}+{\bf G}){\bf r}]$. The
corresponding Dyson equation should be written again to account for the fact
that $\epsilon _{long}$ is now the matrix $\epsilon _{long}({\bf q}+{\bf G},%
{\bf q}+{\bf G}^{\prime }).$ This will reduce the number of iterations even
more.

Finally, we would like to point out that it should be possible to apply the
same idea to the self--consistency problem in the standard band structure
calculation. In the crystal, due to electroneutrality of the charge density, 
$\rho ({\bf G}=0)=0.$ However, for those reciprocal--lattice vectors which
are small, the components of the potential $V({\bf G})\sim \rho ({\bf G})/|%
{\bf G}|^2$ might be large. This is especially the case for large
many--atomic unit cells. As a consequence, the mixing parameter $\lambda
_{mix}\,$ has to be chosen very small. The procedure described above will
require the calculation of the polarizability (\ref{3.26}) with ${\bf q}=0\,$
at each self--consistent iteration, i.e. the response of electrons to the
plane waves $exp[i{\bf Gr}]$ according to the expression (\ref{3.32}) for
all small vectors $|$ ${\bf G|}\leq G_{cutoff}.$ The cutoff can be chosen as
the radius of the smallest first coordination sphere in $G$--space. The
computational time for finding the $\Pi _G({\bf r})$--functions should
presumably not exceed the time of one self--consistent iteration while the
total number of iterations needed to reach the convergency is expected to
decrease by approximately one order of magnitude, which is the case in
linear--response calculation. Note that the idea just outlined is different
from the idea of finding the self--consistent charge transfer in terms of
the linear--response theory \cite{Skriver}. For large cells we are screening
the small $G$--components of the potential which result from some average
density distribution. On the other hand, such details as the charge transfer
between nearest atoms is described by large $G$--components of $\rho ({\bf G}%
).$

\bigskip\ \ 

{\bf IV. APPLICATIONS.}

\smallskip\ 

In recent publications\cite{PHN,EPI,UFN} we have demonstrated the ability of
our linear--response method to compute whole phonon dispersions and
electron--phonon interactions in such complicated systems as transition
metals $Nb$ and $Mo$. In the present paper we will describe application of
the method for calculating phonon dispersions in the materials with a few
atoms per unit cell and with a relatively open crystalline structures. Two
systems have been chosen for the applications. The first one is $Si$ which
is an excellent test case because of its open diamond structure. The second
one is a transition--metal carbide $NbC$. This is a well--known classic
superconductor with $T_c=11.5K$ and its phonon dispersions show many
anomalies that are not present in other simple--metallic and insulating
systems. Studying these anomalies as well as their influence on
superconductivity and transport is interesting in itself and also represents
a hard test for our method. Here we will only describe the calculations for
the phonon dispersion curves in $NbC$ and compare the results with
experiments. The calculated electron--phonon interaction and transport
properties will be published elsewhere.

\smallskip\ 

{\bf a. Si.}

\smallskip\ 

$Si$ is a well studied elemental semiconductor from both experimental and
theoretical sides and its phonon dispersions have been measured long time ago%
\cite{SIEXP}. Recent linear--response\cite{Baroni2,Krak} and supercell\cite
{Chou} calculations have allowed to determine its lattice dynamics for the
wave vectors in the entire Brillouin zone and the results show a good
agreement with the experiment. These calculations were based on the
linear--augmented--plane--wave and plane-wave pseudopotential methods.
Within the localized--orbital representation employed in the LMTO\ method it
is generally difficult to treat the materials with the diamond structure
and, to reach close packing, interstitial spheres are usually placed into
the empty sites of the lattice. This complicates the determination of the
dynamical matrix. However, this problem is avoided in our method by the use
of the Fourier transform for the LMTOs in the interstitial region.

We calculate the dynamical matrix of $Si$ as a function of wave vector for a
set of irreducible ${\bf q}$-points in a $(6,6,6)$--reciprocal lattice grid
(16 points per $1/48$th part of the BZ). The $(I,J,K)$ reciprocal lattice
(or Monkhorst--Pack) grid is defined in a usual manner: ${\bf q}_{ijk}=\frac 
iI{\bf G}_1+\frac jJ{\bf G}_2+\frac kK{\bf G}_3$, where ${\bf G}_1,{\bf G}_2,%
{\bf G}_3$ are the primitive translations in reciprocal space. The details
of the calculations for every ${\bf q}$--point are the following: We use $%
3\kappa -spd$ LMTO basis set (27 orbitals per atom) with the one-center
expansions inside the MT-spheres performed up to $l_{max}=6.$ In the
interstitial region, the $s$, $p$ and $d$ -- basis functions are expanded in
plane waves up to 15.1, 22.3, 31.7 Ry (282, 530, 868 plane waves)
respectively. The induced charge densities and screened potentials are
represented inside the MT--spheres by spherical harmonics up to $l_{max}=6$
and by plane waves with the 110.2 Ry energy cutoff (5208 plane waves) in the
interstitial region. The ${\bf k}$--integration over the BZ is performed
over the $(6,6,6)$ -- grid (the same grid as for the phonon wave vectors $%
{\bf q}$) by means of the improved tetrahedron method\cite{TETR2} which is
identical in the case of $Si$ to the special--point method of Monkhorst and
Pack$.$ The MT--sphere radius was taken to be 2.214 a.u. and the
Barth--Hedin--like exchange--correlation formulae after Ref. %
\onlinecite{Moruzzi} are employed. We use theoretically determined lattice
parameter in the calculation (the volume ratio $V/V_{exp}=0.991$).

Fig.1 shows a comparison between calculated and experimental phonon
dispersion curves along the major high--symmetry directions. The calculated
phonon density of states is plotted at the right part in the figure. The
theoretical frequencies are denoted by circles and the experimental ones are
denoted by triangles. The lines result from the interpolation between the
theoretical points. The calculated and experimental phonon frequencies at
the high-symmetry points $\Gamma $, $X,$ and $L\,$ are also listed in Table
1. We see that the agreement between theory and experiment is very good.
Especially, in the optical region the discrepancy is about 1-1.5\%, which is
surprising because the accuracy of the measured phonon modes is of the same
order of magnitude. We also reproduce the extended flat regions of the
transverse acoustic modes indicating the accurate description of long--range
interactions between $Si$ -atoms as well as the correct long-wavelength
behavior showing the good accuracy of calculated elastic properties of this
crystal. Larger discrepancy is found for the frequencies of the TA modes,
where the theoretical branches are approximately 10\% softer than the
experimental ones. For instance, the calculated frequency of the $X_{TA}$ --
mode is 4.00 THz while $\omega _{exp}(X_{TA})=4.49\pm 0.06$ THz\cite{SIEXP}.
The same kind of discrepancy has also been recently reported in Refs. %
\onlinecite{Krak,Chou}. The agreement is slightly improved when we
recalculate the dynamical matrix at the $X$ -- point using the experimental
lattice constant. We have found that the frequency of the $X_{TA}$ -- mode
is increased from 4.00 to 4.27 THz. This shows that the mode has a large
negative Gruneizen parameter and it is thus very sensitive to the unit--cell
volume used in the calculation. Because of the large LMTO\ basis sets, large 
$l\,_{max}$ and plane--wave energy cutoffs this discrepancy is hard to
relate to the internal parameters in the calculation and it is most likely
connected with the local density approximation.

\smallskip\ 

{\bf b. NbC.}

\smallskip\ 

The lattice dynamical properties of transition-metal carbides and,
especially, $NbC$ have attracted much attention in the past because of the
existence of pronounced anomalies in its acoustic branches and their
influence to superconductivity . While some model calculations of the phonon
dispersions exist in the literature and various mechanisms explaining these
anomalies have been proposed\cite{Sinha}, no {\it ab initio} investigation
of the lattice dynamics for $NbC$ have so far been performed. Here we apply
the linear--response approach to the phonon spectrum of $NbC\,$ in order to
check the accuracy of our method.

The dynamical matrix of $NbC\,$ is calculated at the $29$ irreducible ${\bf q%
}$ -- points of a $(8,8,8)$ reciprocal--lattice grid. The self--consistent
calculations performed for every wave vector involve the following
parameters: $3\kappa -spd$ LMTO basis per $Nb$ atom (27 orbitals) and $%
3\kappa -sp$ LMTO basis per carbon atom (12 orbitals). The one--center
expansions inside the MT--spheres are performed up to $l_{max}=6.$ In the
interstitial region the basis functions are expanded in plane waves up to
13.4, 19.6, 26.9 Ry (136, 228, 338 plane waves) for, respectively, $s,p$ and 
$d$ -- orbitals of $Nb$, and up to 24.1, 35.8 Ry (306, 536 plane waves) for $%
s,p$ --orbitals of $C.$ The changes in the charge densities and the
potentials are represented inside the MT--spheres by spherical harmonics up
to $l_{max}=6$ and by plane waves with an 121 Ry energy cutoff (3382 plane
waves) in the interstitial region. The ${\bf k}$--space integration for the
matrix elements is performed over a $(8,8,8)$ -- grid (the same grid as for
the phonon wave vectors ${\bf q}$) by means of the improved tetrahedron
method\cite{TETR2}. However, the integration weights for the ${\bf k}$%
--points of this grid have been found to take into account the effects
arising from the Fermi surface and the energy bands precisely. This is done
with help of a $(32,32,32)$ grid (897 ${\bf k}$ -- points per $1/48$ BZ) as
we explained in Section III(c) of this paper. The MT--sphere radius of $Nb$
is taken to be 2.411 a.u. and the radius of the carbon sphere is 1.786 a.u.
The Barth--Hedin--like exchange--correlation formulae after Ref. %
\onlinecite{Moruzzi} are employed. As in the case of $Si$, we also use the
theoretically determined lattice parameter in this calculation (the volume
ratio $V/V_{exp}=0.982$).

The results of our calculations are presented in Fig. 2, where we compare
theoretically determined phonon dispersions (circles) with those measured by
inelastic--neutron--scattering technique \cite{NBCEXP} (triangles). The
calculated phonon density of states is plotted at the right part of the
figure. The lines are simply the result of interpolation between the
theoretical points. Since the ${\bf q}-$ grid $(8,8,8)$ considered here is
still too coarse to resolve the anomaly of the longitudinal acoustic branch
near the wave vector $(0.6,0,0)\,2\pi /a\,$, we have performed a separate
calculation for the ${\bf q}$ --point $(0.625,0,0)$ which fits to the $%
(16,16,16)$ -- grid in ${\bf k}$--space. We see that the agreement between
theory and experiment is good. Most of the calculated frequencies agree
within a few percent with those measured despite of the fact that we have
used only 29 ${\bf k}$--points for the BZ--integration. (We list for
comparison our calculated and experimental phonon frequencies at the
high--symmetry points $\Gamma $, $X,$ and $L\,$ in Table 2.) The theory
reproduces the major anomalies presented in the acoustic branches: the
well-known anomaly near the wave vector $(0.6,0,0)\,2\pi /a$ which is also
present and well described\cite{PHN} within our linear--response method in
pure $Nb$ crystal; the anomaly of the longitudinal mode near the wave vector 
$(0.5,0.5,0)\,2\pi /a$ as well as large softening of both TA and LA modes
near the $L$ --point. Moreover, we also predict an anomalous behavior of the
lowest transverse acoustic mode along the $(\xi \xi 0)$ direction near the
wave vector $(0.5,0.5,0)\,2\pi /a$ .Here the frequencies are not known
experimentally. The anomaly found by us is, however, less pronounced
compared to the results of double--shell model calculations of Weber \cite
{Weber} while we have certainly not too many points along this direction to
judge about its exact dispersion.

\bigskip\ 

{\bf V. CONCLUSION.}

\smallskip\ 

In conclusion, we have described in detail an all--electron linear--response
approach based on density functional theory and the LMTO\ technique. The
method is developed to calculate lattice dynamical properties of crystalline
solids and is uniquely applicable for the systems with broad and narrow
energy bands. For test purposes, we have applied the method to compute
phonon dispersions for $Si$ and $NbC$ which have open structures and two
atoms per unit cell. The results of our applications are in a good agreement
with the experiment. We have thus shown that accurate calculations of
lattice dynamics are now possible even for such complicated systems as
transition--metal compounds. In the forthcoming paper\cite{FORTH} we give a
description of our method for calculating electron--phonon interactions and
apply the method to compute lattice--dynamical, superconducting and
transport properties for a large number of elemental metals (a brief report
of this work has appeared already\cite{EPI}). In another publication \cite
{FORTH2} we describe an application of the method for computing
electron--phonon--coupling strength in $Ca-Sr-Cu-O$ high-$T_c$
superconductor.

\bigskip\ 

{\bf Acknowledgments:} The author is indebted to Professor O. K. Andersen
for many helpful discussions and to Dr. O. Jepsen for careful reading of the
manuscript prior to the publication. The work was partially supported by
INTAS (93-2154), ISF(MF-8300), and RFFI grants.

\bigskip\ 

{\bf APPENDIX: Fourier transform of pseudoLMTOs.}

\smallskip\ 

Consider a Hankel function $K_{\kappa L}({\bf r})=K_{\kappa l}(r)i^lY_{lm}(%
{\bf r})$ of energy $\kappa ^2$ which is singular at the origin. The
three--dimensional Fourier transform of this function $K_{\kappa L}({\bf k})$
is known to behave as $k^{l-2}$ for large $k.\,$The task is to substitute
the divergent part of $K_{\kappa l}(r)\,$ inside some sphere $s$ by a smooth
regular but otherwise arbitrary function. This function is chosen so that
the Fourier transform is convergent fast. In the full--potential LMTO method
of Ref. \onlinecite{Weyrich}, the augmenting function is the linear
combination of the Bessel function $J_{\kappa L}$ and its energy derivative $%
\dot J_{\kappa L}$ matched together with its first--order radial derivative
with the Hankel function at the sphere boundary. The Fourier transform
becomes convergent as $k^{-4}$. One can obviously include higher--order
energy derivatives $\stackrel{(n)}{J}_{\kappa L}$ in order to have a smooth
matching at the sphere up to the order $n$. This was done in connection with
the problem of solving the Poisson equation in Ref. \onlinecite{Weinert}.
The Fourier transform here converges as $k^{-(3+n)}$ but the prefactor
increases as $(2l+2n+3)!!$ and this prohibits the use of large values of $n$%
. A similar procedure has been also used in the LMTO\ method of Ref. %
\onlinecite{Wills}. In the present work we will use a different approach
based on the Ewald method. Instead of substituting the divergent part only
for $r<s$ we consider the solution of the equation:

\begin{equation}
(-\nabla ^2-\kappa ^2)\tilde K_{\kappa L}({\bf r})=a_l\left( \frac rs\right)
^le^{-r^2\eta ^2+\kappa ^2/\eta ^2}i^lY_{lm}({\bf r}),  \label{a.1}
\end{equation}
The function on the right-hand side of the Helmholtz equation is a decaying
Gaussian. The parameter $a_l$ is a normalization constant: $a_l=\sqrt{2/\pi }%
(2\eta ^2)^{l+3/2}s^{2l+1}/(2l-1)!!$. The most important parameter is $\eta $%
. It is chosen such that the Gaussian is approximately zero when $r>s$ and $%
\eta $ must depend on $l$ as well as the sphere radius $s.$ The solution $%
\tilde K_{\kappa L}({\bf r})$ is thus the Hankel function for large $r$, it
is a regular function for small $r$ and it is smooth together with its
radial derivatives at any $r$. The function $\tilde K_{\kappa l}(r)$ can be
calculated in terms of the following error-function-like contour integral:

\begin{equation}
\tilde K_{\kappa l}(r)=\frac{(2s)^{l+1}}{\sqrt{\pi }(2l-1)!!}%
r^l\int_{0+}^\eta \xi ^{2l}e^{-r^2\xi ^2+\kappa ^2/4\xi ^2}d\xi .
\label{a.2}
\end{equation}
When $\eta \rightarrow \infty $ this integral is known as the Hankel
integral. The most important result is that the Fourier transform of $\tilde 
K_{\kappa l}(r)$ decays exponentially. It is given by:

\begin{equation}
\tilde K_{\kappa l}(r)=\frac 2\pi \frac{s^{l+1}}{(2l-1)!!}\int_0^\infty
k^2dkj_l(kr)\frac{k^le^{(\kappa ^2-k^2)/4\eta ^2}}{k^2-\kappa ^2}.
\label{a.3}
\end{equation}
Restoring the original notations, the pseudoLMTOs $\tilde \chi _{\kappa RL}^{%
{\bf k}}({\bf r})$ are the Bloch waves of wave vector ${\bf k}$ as defined
in Eq. (\ref{3.9}). The Fourier coefficients $\tilde \chi _{\kappa RL}({\bf k%
}+{\bf G})$ are given by:

\begin{eqnarray}
&&\tilde \chi _{\kappa RL}({\bf k}+{\bf G})=\frac{4\pi }{\Omega _c}\frac{%
s_R^{l+1}}{(2l-1)!!}\frac{|{\bf k}+{\bf G}|^l}{|{\bf k}+{\bf G}|^2-\kappa ^2}%
\times  \nonumber \\
&&e^{(\kappa ^2-|{\bf k}+{\bf G}|^2)/4\eta _{Rl}^2}Y_L({\bf k}+{\bf G})e^{-i(%
{\bf k}+{\bf G}){\bf R}},  \label{a.4}
\end{eqnarray}
where $\Omega _c$ is the volume of the unit cell and where we have
subscripted $\eta $ with the indexes $Rl$ and $s$ with $R.$

In practical calculations the parameter $\eta _{Rl}$ can be chosen from the
ratio between the Hankel function at the sphere and the solution of Eq. (\ref
{a.1}), i.e. $\tilde K_{\kappa l}(s_R)/K_{\kappa l}(s_R)=1+\delta .$ The
error $|\delta |$ is usually taken not larger than $0.03$ which leads to the
number of plane waves per atom needed for the convergency in (\ref{3.9})
varying from 150 to 250 when $l=2.$ For the $s,p-$orbitals this number is
smaller by a factor of $2-3.$

\end{multicols}

\bigskip\ 

\begin{table}
\caption{Comparison between calculated and experimental phonon frequencies
at the high-symmetry points $\Gamma $, $X$, and $L$ for $Si$ [THz]. } 
\begin{tabular}{lllllllll}
& $\Gamma _{LTO}$ & $X_{TA}$ & $X_{LAO}$ & $X_{TO}$ & $L_{TA}$ & $L_{LA}$ & $%
L_{TO}$ & $L_{LO}$ \\ \hline
theory & 15.56 & 4.00 & 12.27 & 13.90 & 3.09 & 11.20 & 14.78 & 12.38 \\ 
exp$^a$. & 15.53 & 4.49 & 12.32 & 13.90 & 3.43 & 11.35 & 14.68 & 12.60 \\ 
&  &  &  &  &  &  &  & 
\end{tabular}
$^a$Reference \onlinecite{SIEXP} 
\end{table}

\bigskip\ 

\begin{table}
\caption{Comparison between calculated and experimental phonon frequencies
at the high-symmetry points $\Gamma $, $X$, and $L$ for $NbC$ [THz]. } 
\begin{tabular}{llllllllll}
& $\Gamma _{LTO}$ & $X_{TA}$ & $X_{LA}$ & $X_{TO}$ & $X_{LO}$ & $L_{TA}$ & $%
L_{LA}$ & $L_{TO}$ & $L_{LO}$ \\ \hline
theory & 17.05 & 6.37 & 7.51 & 17.64 & 18.65 & 4.26 & 6.02 & 18.82 & 21.60
\\ 
exp$^a$. & 16.70 & 6.35 & 7.30 & 17.20 & 17.80 & 4.00 & 6.00 & -- & 19.20 \\ 
&  &  &  &  &  &  &  &  & 
\end{tabular}
$^a$Reference \onlinecite{NBCEXP} 
\end{table}

\bigskip\ 

Fig. 1. Calculated phonon dispersion for Si (circles) along the
high-symmetry directions in comparison with the experiment \cite{SIEXP}
(triangles). The lines are the result of interpolation between theoretical
points. Also shown is the calculated phonon density of states (DOS).

Fig. 2. Calculated phonon dispersion for NbC (circles) along the
high-symmetry directions in comparison with the experiment \cite{NBCEXP}
(triangles). The lines are the result of interpolation between theoretical
points. Also shown is the calculated phonon density of states (DOS).

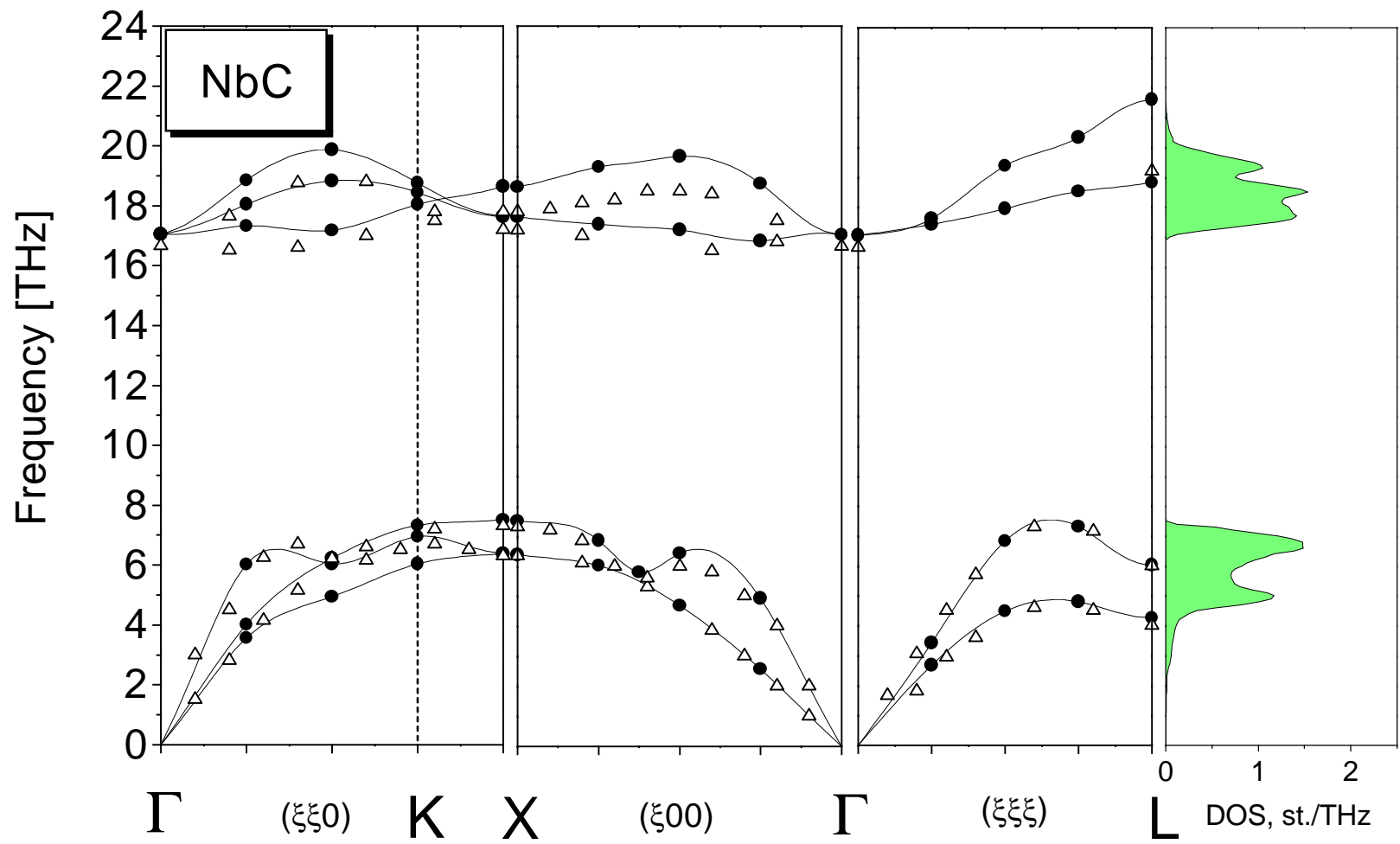
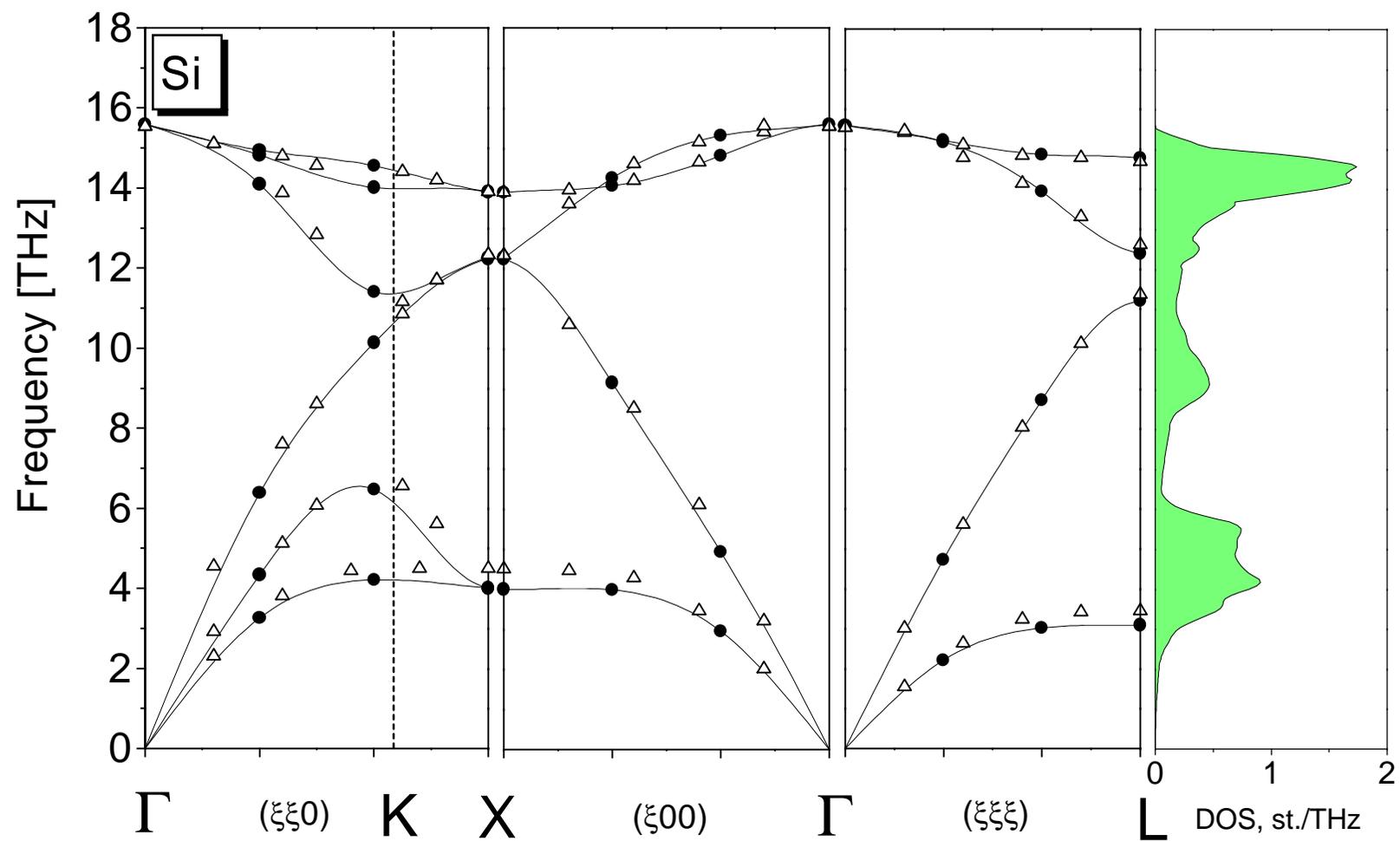
\end{document}